\documentclass[fleqn,10pt]{wlscirep}

\usepackage[utf8]{inputenc}
\usepackage[T1]{fontenc}
\usepackage{amsmath,amssymb,bm, bbold, physics}% bold math
\usepackage{float}%for using [H] to force fig to stay in its place

\usepackage{lipsum}
\usepackage{lmodern}
\usepackage[most]{tcolorbox}
\usepackage{wrapfig}
\usepackage{comment}

\newcommand{\kbm}{\bm{k}}

\newcommand{\Rmnum}[1]{\expandafter\@slowromancap\romannumeral #1@}

\newcommand{\moire}{moir\'e }
\newcommand{\Moire}{Moir\'e }

\newcommand{\Vxxw}{$V_{xx}^\omega$}
\newcommand{\Exxw}{$E_{xx}^\omega$}

\newcommand{\Vxytw}{$V_{xy}^{2\omega}$}
\newcommand{\Exytw}{$E_{xy}^{2\omega}$}

\newcommand{\Vbynm}{V\,nm$^{-1}$}
\newcommand{\D}{$D/\epsilon_0$}

\newcommand{\Vxytwnorm}{$\frac{V_{xy}^{2\omega}}{(V_{xx}^{\omega})^2}$}

\newcommand{\Rxx}{$R_{xx}$}
\newcommand{\Rxy}{$R_{xy}$}

\newcommand{\sxxsq}{$\sigma_{xx}^2$}

\newcommand{\uvec}[1]{\boldsymbol{\hat{\text{#1}}}}

\title{
Tunable \moire materials for probing Berry physics and topology}

\author[1,2$\dagger$]{Pratap Chandra Adak}
\author[1$\dagger$]{Subhajit Sinha}
\author[3]{Amit Agarwal}
\author[1,*]{Mandar M. Deshmukh}
\affil[1]{Department of Condensed Matter Physics and Materials Science, Tata Institute of Fundamental Research, Mumbai, India}
\affil[2]{Department of Physics, City College of New York, New York, USA}
\affil[3]{Department of Physics, Indian Institute of Technology, Kanpur, India
}
\affil[$\dagger$]{These two authors contributed equally to this work}
\affil[*]{e-mail: deshmukh@tifr.res.in}

\begin{abstract}
Berry \textcolor{black}{curvature} physics \textcolor{black}{and quantum geometric effects} have been instrumental in advancing topological condensed matter physics in recent decades. Although Landau level-based flat bands and conventional 3D solids have been pivotal in exploring rich topological phenomena, they are constrained by their limited ability to undergo dynamic tuning. In stark contrast, moiré systems have risen as a versatile platform for engineering bands and manipulating the distribution of Berry curvature in momentum space. These moiré systems not only harbor tunable topological bands, modifiable through a plethora of parameters, but also provide unprecedented access to large length scales and low energy scales. Furthermore, they offer unique opportunities stemming from the symmetry-breaking mechanisms and electron correlations associated with the underlying flat bands that are beyond the reach of conventional crystalline solids. A diverse array of tools, encompassing quantum electron transport in both linear and non-linear response regimes and optical excitation techniques, provide direct avenues for investigating Berry physics. This review navigates the evolving landscape of tunable moiré materials, highlighting recent experimental breakthroughs in the field of topological physics. Additionally, we delineate several challenges and offer insights into promising avenues for future research.
\end{abstract}

\begin{document}
\renewcommand{\hbar}{\mathchar'26\mkern-9mu h}%SS-strangely \hbar was not working until I defined this newcommand

\flushbottom
\maketitle

\thispagestyle{empty}

\section{Introduction}

Materials have played a pivotal role in advancing science and technology, as exemplified many times over the last century. 
A prime illustration is the development of semiconductors, which underpin our modern technological landscape. 
Materials discovery is crucial to the development of new technologies. 
A rational rather than a combinatorial approach to material discovery requires a system whose parameters can be tuned to understand and optimize the properties of the system \cite{kennes_moire_2021}.  
One such versatile and adaptable platform for material discovery and fundamental physics is the \moire superlattice\cite{he_moire_2021}, arising when layers of 2D materials with the same, or different, lattice constants are twisted or stacked.

2D van der Waals materials with a huge library of exfoliable materials \cite{mounet_two-dimensional_2018} have led to complex heterostructures \cite{geim_van_2013, novoselv2018vdWheterostructures} for the exploration of a wide gamut of physics and novel device architectures \cite{liu_van_2016}. 
The ability to manipulate electronic properties by varying the twist angle, or due to lattice mismatch, in the heterostructure adds a crucial new ingredient not found in native 2D materials. 
The additional periodicity with a large superlattice length scale ($l$) plays an important role in sculpting electronic wavefunctions with novel properties, as the wavefunctions extend over a large number of atoms. 
In the momentum space picture, bands are folded within a mini Brillouin zone endowed with numerous band crossings. 
Interlayer hybridization, coupled with lattice relaxation, gives rise to mini bands, often exhibiting minimal dispersion that allows electron correlation to dominate over the kinetic energy. 
Notably, the recent surge of interest in \moire materials has been ignited by the emergence of remarkable correlation phenomena\cite{andrei_graphene_2020, andrei_marvels_2021}, including Mott insulators\cite{cao_correlated_2018}, superconductivity\cite{cao2018unconventional}, and magnetism\cite{serlin2020intrinsic, sharpe_emergent_2019} observed in magic-angle twisted bilayer graphene (TBG) and subsequently in other \moire systems.

Long before the advent of 2D materials, spanning several decades, disparate scientific disciplines of magnetism had laid the foundations for the emergence of topological condensed matter physics, spurred by the quest to understand anomalous Hall effect \cite{nagaosa_anomalous_2010}, and quantum Hall \cite{von_klitzing_quantized_1986} physics.
The intriguing ideas of the geometric, or the Berry phase in condensed matter and its associated gauge field - Berry curvature, are at the heart of new insights about the topology of electronic states in lattices\cite{xiao_berry_2010}. 
In quantum materials, the Berry curvature plays a central role in several phenomena, such as the anomalous Hall effect, valley Hall effect, quantum spin Hall effect, and the concept of topological invariants. 
Furthermore, other band geometric quantities, such as the quantum metric, the orbital magnetic moment, and the Berry curvature dipole, also induce interesting Hall responses in transport and optics \cite{gao_quantum_2023,wang_quantum_2023,liu_quantum_2019,sodemann_PRL2015_quant,sinha_NP2022_berry,bhalla_PRL2022_reso}.
\textcolor{black}{Quantum geometry is a broader term that encompasses Berry curvature, quantum
metric, and other higher-order tensors that depend on the electron wavefunction~\cite{ahn_riemannian_2022}}.

\begin{figure}[H]
	\centering
	\includegraphics[width=17cm]{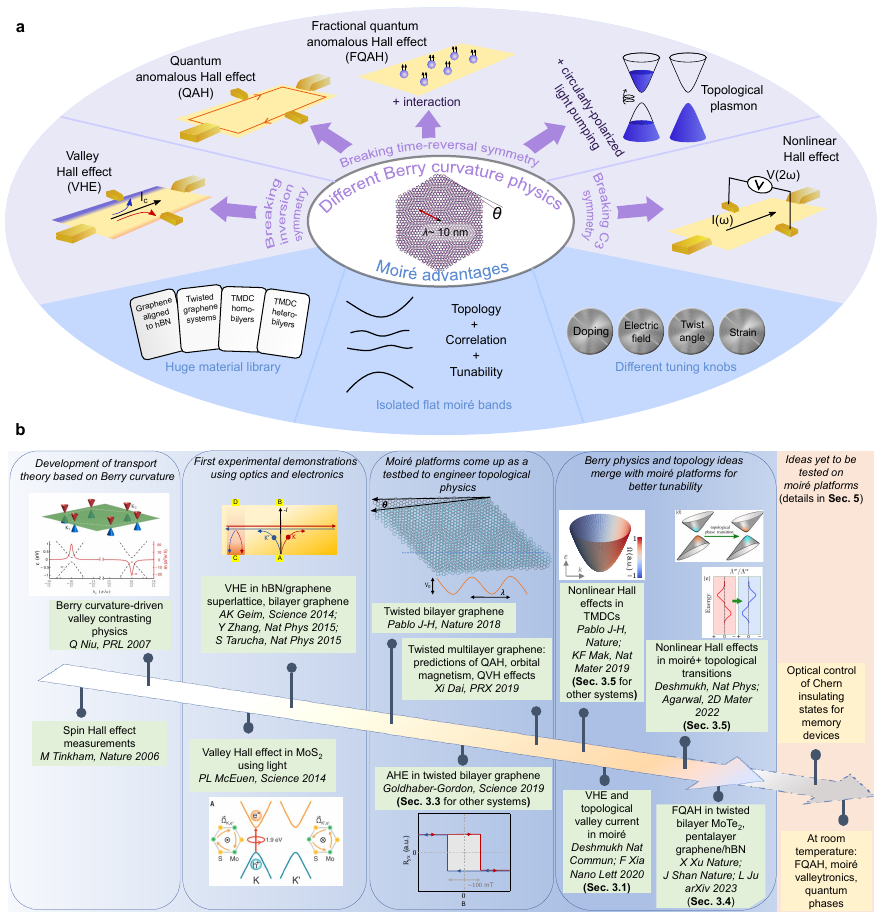}
	\caption[xxx]{\label{fig:timeline}{\textbf{Historical development of Berry physics in \moire systems.}
    \textbf{a,}~A diagram showing advantages of \moire materials and showcasing various Berry physics explored therein.
    \textbf{b,}~Timeline of the development of Berry physics.
    The list of Refs. included are (\cite{valenzuela2006direct,xiao_valley-contrasting_2007, mak_valley_2014, gorbachev_S2014_detect, sui_NP2015_gate, shimazaki_NP2015_gen, cao2018unconventional, sharpe_emergent_2019, liu_quantum_2019, ma_observation_2019, kang_nonlinear_2019, sinha_NC2020_bulk, ma_moire_2020, sinha_NP2022_berry, chakraborty2022nonlinear, 
    cai_signatures_2023, zeng2023thermodynamic, lu_fractional_2023}). VHE: valley Hall effect, QAH: quantum anomalous Hall, QVH: quantum valley Hall, AHE: anomalous Hall effect, FQAH: fractional quantum anomalous Hall effect.
    \textcolor{black}{In panel \textbf{b}, the images in the first box, the lower panel of the second box, and the fourth box (right upper and right lower) are reproduced from Ref.~\cite{xiao_valley-contrasting_2007} (APS), Ref.~\cite{mak_valley_2014} (AAAS), and Ref.~\cite{chakraborty2022nonlinear} (IOP) respectively.}
    }}
\end{figure}

More recently, two-dimensional \moire materials have opened a new avenue for exploring Berry physics, with their large and adjustable length scales enabled by modulating the twist angle and tunable flat electronic bands with energy scales varying from a few meV to 100s of meV. 
The tunability of the electronic structure in \moire materials has been used for several novel transport and topological phenomena, including the anomalous Hall effect, valley Hall effect, nonlinear Hall effect, and orbital magnetism; see Fig.~\ref{fig:timeline}a for the range of experimental knobs and the resulting physical systems realized with \moire systems. 
Notably, \moire materials combine the two interesting facets of momentum space topology and strong correlation physics within a single platform, a rarity in conventional atomic crystals. We present a timeline of key discoveries in 2D and \moire  materials related to topological and Berry physics through transport and optics experiments 
in Fig.~\ref{fig:timeline}b.

For researchers in this rapidly evolving field, there are comprehensive reviews of Berry physics \cite{chang_colloquium_2023,AHE_MacDonald_2002,xiao_berry_2010} in the broader materials context and those specifically centered on 2D materials \cite{ren_topological_2016} together with excellent pedagogical sources \cite{vanderbilt2018berry,bernevig2013topological}.
In this review, we focus on 2D \moire materials to explore the interplay of their tunable electronic properties and Berry curvature and how this interplay manifests in electronic transport phenomena. 
We aim to summarize key experimental findings and present exciting prospects on the horizon, stemming from the confluence of topology and correlations in \moire materials.

\section{Opportunities provided by versatile \moire platform} 

Over the past two decades, Berry physics has been probed across a wide range of systems, significantly enriching our understanding. However, a platform that serves as a quantum simulator \cite{kennes_moire_2021}, a system with many tunable parameters, for topological studies is desirable. As we elucidate next, \moire materials bring unique advantages to the exploration of Berry physics. 

\subsection{A simple approach to generate topological bands}

\begin{figure}
	\centering
	\includegraphics[width=16cm]{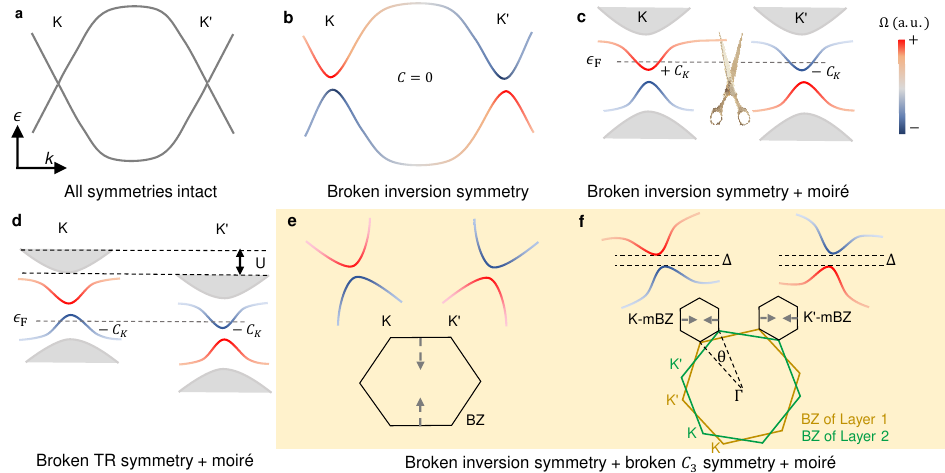}
	\caption[xxx]{\label{fig:knobs}{\textbf{Breaking different symmetries in \moire superlattices.}
    \textbf{a,}~Schematic bandstructure of graphene close to K and K' valley.
    \textbf{b,}~Schematic bandstructure of gapped monolayer graphene, where the broken inversion symmetry introduces a gap. The color indicates $\Omega$.
    As $\Omega$ is equal and opposite for the two valleys $K$ and $K'$, the Chern number of each band is zero.
    \textbf{c,}~Schematic bandstructure for a \moire superlattice with broken inversion symmetry. Due to the \moire gaps, the $K$ and the $K'$ bands decouple. This allows a non-zero valley Chern number $C_K$($C_{K'}$) to be defined at each valley $K$($K'$). The pair of scissors is just a schematic representation of the decoupling of the $K$ and $K'$ valleys.
    %Regardless of where the Fermi level $\epsilon_F$ is placed, there is no net edge state as time-reversal symmetry guarantees $C_{K'}=-C_K$.
    \textbf{d,}~The band structure of a \moire superlattice close to $K$ (left) and $K'$ (right) valley in the absence of time-reversal (TR) symmetry. A broken TR symmetry (due to electronic interaction or external magnetic field) lifts the $K$ and $K'$ degeneracy.
    The flat band's red (blue) color represents positive (negative) $\Omega$.
    This lifting of the degeneracy gives rise to a net edge state and the anomalous Hall effect. Here, U represents the interaction-induced exchange.
    \textbf{e,} Schematic showing the effect of breaking C$_3$ symmetry (via strain) on the band structure. Strain tilts the low-energy band structure in a gapped Dirac system.
    Breaking inversion and C$_3$ symmetries together, keeping TR symmetry intact, creates Berry curvature dipole (BCD) (further explained in theory box).
    The arrows indicate the distortion of the Brillouin zone (BZ) due to strain.
    \textbf{f,} Strain modifies the band structure of the \moire systems in the K and K' \moire Brillouin zone (mBZ).
    \moire periodicity enhances the BCD.
    Here, for simplicity, the \moire unit cells are not distorted in the schematic.
    }}
\end{figure}

A key ingredient for topological materials is non-zero Berry curvature $(\Omega)$, which originates from the quantum mechanical wavefunction of the Bloch quasiparticles.
The existence of non-zero Berry curvature necessitates the breaking of either the inversion or time-reversal symmetry (see theory box).  
Moiré systems are often realized by creating a superlattice that allows for the breaking of inversion symmetry while preserving time-reversal symmetry. 
As an illustration, graphene, the archetypal 2D material with a hexagonal lattice structure, has zero Berry curvature for both valence and conduction bands that touch at two degenerate valleys, $K$ and $K'$ (Fig.~\ref{fig:knobs}a).
However, when brought into near alignment with hexagonal boron nitride (hBN), a lattice with a closely matched lattice constant,  the sublattice symmetry is broken; this results in the formation of an energy gap\cite{yankowitz_emergence_2012} and a finite valley contrasting Berry curvature\cite{xiao_valley-contrasting_2007}, as exemplified in Fig.~\ref{fig:knobs}b.
In Bernal-stacked multilayer graphene, a perpendicular electric field breaks the inversion symmetry, yielding non-zero Berry curvature.
The strength of the Berry curvature is inversely related to the magnitude of the gap,  offering a tunable parameter\cite{sinha_NC2020_bulk}. 
However, merely breaking inversion symmetry can still leave the two valleys of graphene-like systems connected at high energies, making the valley-specific Chern numbers ill-defined\cite{PNAS_song_topological_2015}.  In this scenario, the Chern number of the overall band remains zero as the Berry curvature around the two valleys is related through time-reversal symmetry.

\Moire superlattice provides a pivotal pathway to harness nontrivial topological invariants by "dicing" the electronic bands in energy via the Bragg scattering of electrons arising from the new \moire periodicity\cite{PNAS_song_topological_2015, wolf_substrate-induced_2018}. 
The \moire periodicity $(l)$ is given by $l=\frac{(1+\delta)a}{\sqrt{2(1+\delta)(1-\cos{\theta})+\delta^2}}$, where $\delta$ is the fractional lattice mismatch, $a$ is the lattice constant, and $\theta$ is the relative rotation between the 2D layers, commonly referred as the twist angle\cite{yankowitz_emergence_2012}. 
The low-energy physics of \moire superlattices is understood in the reduced \moire Brillouin zone (mBZ) with mini bands resulting from band folding. 
As the bands are "diced", \moire bands associated with the two valleys become isolated, as depicted in Fig.~\ref{fig:knobs}c, where the scissors denote the isolation of bands from each other. 
This band reconstruction has profound implications, enabling the \moire bands for each of the two valleys to possess unique valley Chern numbers, $C_v = (1/2\pi) \int_\text{mBZ}~\Omega(\bm{k})~d^2k$.
While time reversal symmetry ensures $C_K+C_{K'}=0$, a valley-dependent exchange term can break the symmetry between the two \moire valleys (as depicted in the Fig.~\ref{fig:knobs}d) and lead to a host of interesting Hall effects that we discuss later.

Non-zero valley Chern number, $C_v=\pm 1$ was proposed in graphene aligned to hBN for commensurate stacking, whereas $C_v = 0$ for incommensurate stacking\cite{PNAS_song_topological_2015}.
This suggests a crucial role of the twist angle in tuning the topological properties in \moire superlattice.
Later, many other \moire superlattices possessing Chern bands\cite{zhang_nearly_2019,chittari_gate-tunable_2019, liu_quantum_2019 } have been proposed or realized along with the availability of different knobs to tune the topology, as we discuss next.

\subsection{Many knobs for engineering symmetries and topological properties} 

\begin{figure}
	\centering
	\includegraphics[width=13cm]{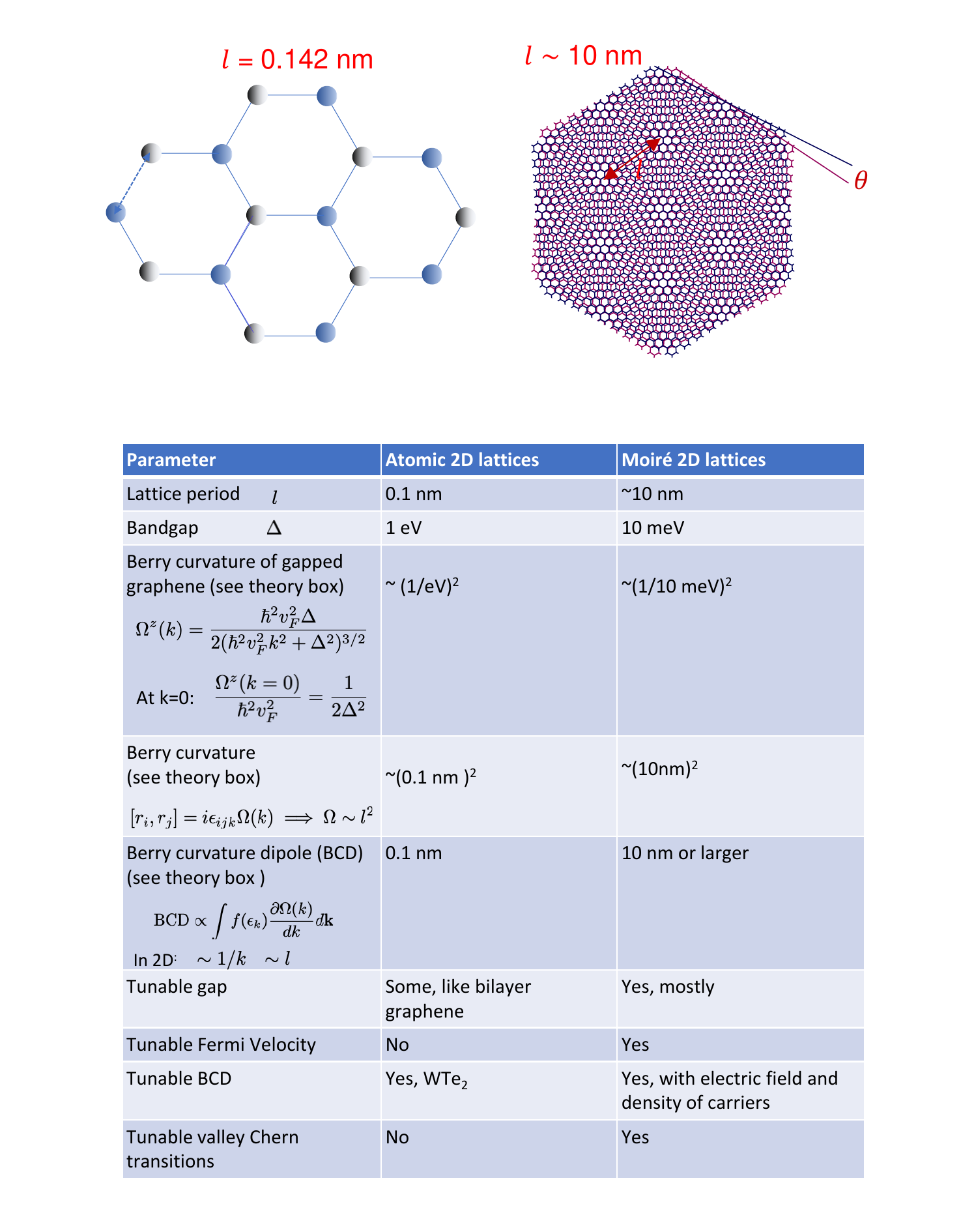}
	\caption[xxx]{\label{fig:comparisonwithmoire}{\textbf{What \moire superlattices add to the exploration of Berry physics beyond atomic lattices.} 
  (top) The atomic lattice and the \moire lattice and their lattice constants. (bottom) %This panel 
  Comparison of different aspects of atomic lattices with \moire superlattices. % and the scaling of with the key length and energy scales. 
  The high tunability of some properties in \moire superlattices highlights its advantages. 
  The scaling with lattice length uses intuitive arguments to develop physical insight.
  \textcolor{black}{The numbers indicate the order of magnitude.}}}
\end{figure}

An avenue for in-situ tuning common to a broad spectrum of 2D materials is doping charge carriers via an electrostatic gate, thereby enabling precise control over the Fermi energy.
Moiré systems, stemming from their constituent 2D materials, inherit this capability of Fermi energy tuning through electrostatic gating.
What further sets \moire systems apart is their characteristic small energy gaps, typically around 10 meV, and narrow bandwidth, approximately 10 meV, making them amenable to exploration over a significant range of energies to sample the energy-dependent Berry physics (see Fig.~\ref{fig:comparisonwithmoire}).
In particular, the large lattice constant of \moire superlattice plays a vital role by allowing for the manipulation of a much larger number of electrons per lattice through electrostatic gating. 
In other words, electrostatic gating can be employed to occupy multiple \moire bands, thereby enabling direct experimental access to the Chern numbers of numerous individual bands. 
This capability is a significant departure from conventional 2D lattices, where filling a complete band and accessing the Chern number of successive bands is often impractical due to the requirement to traverse the entire Brillouin zone. 

Introducing a perpendicular electric field through dual gates is instrumental in breaking inversion symmetry, a phenomenon we have previously emphasized.
Intriguingly, the perpendicular electric field can dynamically tune the valley Chern number in several \moire systems\cite{zhang_spin-polarized_2022}.
For example, in ABC trilayer graphene aligned to hBN, reversing the electric field can switch the Chern number from zero to a finite value, enabling the fascinating prospect of electric field-tunable ferromagnetism\cite{chen_tunable_2020}.
Other notable systems to observe electric field tunable topological physics\cite{zhu_voltage-controlled_2020} include twisted mono-bilayer graphene\cite{polshyn2020electrical}, twisted double-bilayer graphene (TDBG)\cite{sinha_NP2022_berry, adak_perpendicular_2022}, twisted trilayer graphene\cite{hao_electric_2021}, etc.

In addition to electrostatic gating, the twist angle plays a paramount role in controlling interlayer hybridization, thereby modulating electronic properties in \moire systems. 
A profound example is the emergence of flat bands in twisted graphene only at certain angles, referred to as the magic angles. 
Furthermore, the twist angle also serves as a knob for tuning the Chern number of \moire bands\cite{zhang_nearly_2019}.
Twisted architecture also offers another way to design topology besides controlling the small twist angle. 
For instance, in TDBG it is predicted that Chern numbers are distinct when the twist angle is $\theta$ or $180-\theta$ \cite{crosse_hofstadter_2020,koshino_band_2019,chebrolu_flat_2019,liu_higher-order_2021}. \textcolor{black}{However, we note that a recent experiment found similar symmetry-broken Chern insulator states in both configurations~\cite{he_symmetry-broken_2023}.}
Nevertheless, novel topological architectures in chiral stacked \moire structures using graphene and transition metal dichalcogenide (TMDC) have been predicted to lead to exciting physics\cite{mannai_stacking-induced_2023, PhysRevLett.120.046801}. 

\Moire systems exhibit additional symmetries, such as $C_3$ symmetry, which can be broken intentionally or unintentionally due to strain introduced during the fabrication of devices\cite{lau_reproducibility_2022}. 
The $C_3$ symmetry is typically broken inadvertently because of the non-uniform distribution of strain in the system due to the stacking process where little control over strain exists. 
Figures~\ref{fig:knobs}e and \ref{fig:knobs}f illustrate the consequence of strain -- it effectively tilts the energy bands, resulting in an asymmetric distribution of Berry curvature in the \moire BZ (see theory box). 
Consequently, breaking $C_3$ symmetry 
\textcolor{black}{in 2D materials leads to a non-zero first moment of Berry curvature, the Berry curvature dipole (BCD)}
\cite{hu_nonlinear_2022,zhang_giant_2022,chakraborty2022nonlinear}. 
The nonzero BCD generates a non-linear response, like the non-linear Hall effect.
Moreover, BCD can serve as a sensitive indicator of topological transitions, an aspect we discuss later in detail.

Van der Waals materials with ferroelectric \cite{zhang_ferroelectric_2023}, and flexoelectric coupling  \cite{mcgilly_visualization_2020} offer additional ways to tune the band structure of \moire systems.
The flexoelectric mechanism, where the strain gradient leads to electric polarization \cite{kalinin_electronic_2008,wang_flexoelectricity_2019,weston_interfacial_2022}
provides a way to in-situ engineer the band structure, hence the Berry curvature distribution.
Such tuning can lead to strain-induced valley Hall or non-linear Hall response \cite{kang_switchable_2023, sinha_NP2022_berry}.

One of the crucial advantages of the van der Waals heterostructures is the ability to stack materials with very different properties. This opens up the possibility of using proximity with a ferromagnet to induce exchange in one of the spin-bands and spin-orbit interaction in the \moire bands. The coupling between spin and valley degrees can lead to effects of Berry curvature-based control of spin \cite{zhong_van_2017}. 
In addition, several studies have used WSe$_2$ to stabilize the \moire structures, enhancing superconducting response \cite{lin_zero-field_2022,arora_superconductivity_2020}. 
The exact microscopic role of WSe$_2$ is not fully understood, but several works have reported the ability to tune superconductivity and ferromagnetism \cite{lin_spin-orbitdriven_2022} in \moire systems. Proximity-induced effect due to exchange and spin-orbit interaction in \moire flat band systems is an area where future studies could lead to novel physics and devices. 

\subsection{Large \moire materials library} 
  
One of the key advantages of the \moire platform is a gamut of materials \cite{mounet_two-dimensional_2018} available for engineering the electronic bands and the facile tuning of the effective Hamiltonian using the twist angles.
Monolayer graphene and few-layer graphene offer a platform to start the engineering of the Dirac bands. 
The transition metal dichalcogenides (TMDC) are the simplest realization of massive Dirac fermions with Berry curvature at the band edges. 
\textcolor{black}{While the large band gaps in TMDCs make the intrinsic Berry curvature small, \moire materials with TMDCs do not have this constraint. 
Furthermore, TMDCs}
are renowned for the strong spin-orbit interaction that allows the coupling of spin and orbital physics. 
Since the spin-orbit interaction strength varies across the family, a suitable choice can be made for tailoring the electronic bands.

As the number of \moire systems employing different graphene or TMDC materials is ever expanding, another exciting avenue is \moire-of-\moire \cite{uri_superconductivity_2023}.
Berry curvature depends on the \moire structure and associated superlattice length scale that governs low-energy physics. 
While real atomic lattices have a length scale of $\sim$0.1~nm, \moire lattices have a length scale of $\sim$10~nm, and metamaterials created using \moire of \moire can have length scales of $\sim$100~nm. 
Many aspects of Berry physics can be enhanced by tuning the unit cell of the lattice (see theory box and Fig.~\ref{fig:comparisonwithmoire}). 
Tuning the lengthscale using \moire or \moire-of-\moire offers the opportunity to study a wide range of Berry curvature and its higher-order moments.

\subsection{Interplay of electronic interactions and topology} 

A unique facet of \moire materials is the combination of rich Berry curvature structure and flat bands that host correlations. The twist angle and the vertical electric field in some materials allow tuning of the Fermi velocity. The control of the flatness of bands, in turn, decides the emergence of symmetry-broken correlated states.
\textcolor{black}{As we discuss later, \moire-induced exchange energy in the spin or valley sector can be comparable to the energy separation of the relevant energy levels leading to spontaneous symmetry-breaking.}
With the wide number of materials available for making vdW heterostructure, one also has indirect control of exchange energies by engineering the screening environment of the moiré.  

Prominent examples where correlations and \moire quantum geometry come together are the QAH (quantum anomalous Hall effect) /CI (Chern Insulator) and FQAH (fractional quantum anomalous Hall effect) /FCI  (fractional Chern Insulator)\cite{zhang_nearly_2019, wu_topological_2019_2, crepel_anomalous_2023}. 
The spontaneous breaking of symmetry in flat bands gives rise to non-zero Chern numbers that are a hallmark of QAH/CI and FQAH/FCI -- an aspect we discuss in detail later.  Some of the consequences of interactions can be extended to charge-neutral excitations where excitonic topological bands have been predicted\cite{wu_topological_2017}; recent demonstration of an excitonic correlated insulator \cite{xiong_correlated_2023} open up many possibilities in this direction. In addition, recent theoretical ideas about the connection of flat band superconductivity with the quantum metric provide a glimpse of intriguing possibilities. 

Having discussed some of the advantages that \moire materials bring to the exploration of topological condensed matter physics, Fig.~\ref{fig:comparisonwithmoire} succinctly summarizes the key advantages of the \moire superlattice over atomic lattices.

\section{Experiments on Berry curvature physics in \moire materials} 
We now discuss some key experiments probing the Berry physics of \moire platform.
\textcolor{black}{We note that strong correlations in \moire{} superlattices give rise to a plethora of interaction-driven phenomena~\cite{cao_correlated_2018, lu_superconductors_2019, tang_simulation_2020, xu_correlated_2020, shimazaki_strongly_2020, choi_correlation-driven_2021, JiaLi_Coulomb_screening_2021, burg_emergence_2022} such as correlated insulators, superconductivity, etc., discussed at length in recent reviews~\cite{andrei_graphene_2020, balents_superconductivity_2020, mak_semiconductor_2022, bhowmik_emergent_2023}.
Our focus centers on phenomena where topology manifests more explicitly.
We start with valley Hall physics and its quantized counterpart, which are not exclusive to \moire systems, nevertheless provide the early indication of topological characters of \moire bands.
Furthermore, the tunability of \moire materials provides new opportunities, such as controlling the quantum geometric structure of the bands by tuning the bandwidth using the electric field. 
Then we discuss both integer and fractional quantum anomalous Hall effects.
Our primary emphasis is on studies in the absence of a magnetic field, even though there are noteworthy studies of Chern insulator states and fractional Chern insulator states that require nonzero magnetic fields. 
We further review recent developments in \moire materials beyond the linear Hall responses -- nonlinear Hall effect, Berry plasmons, and quantum metric-induced phenomena.
}
 
\subsection{Valley Hall Effect (VHE) and topological valley current}

\begin{figure}[H]
	\centering
	\includegraphics[width=18cm]{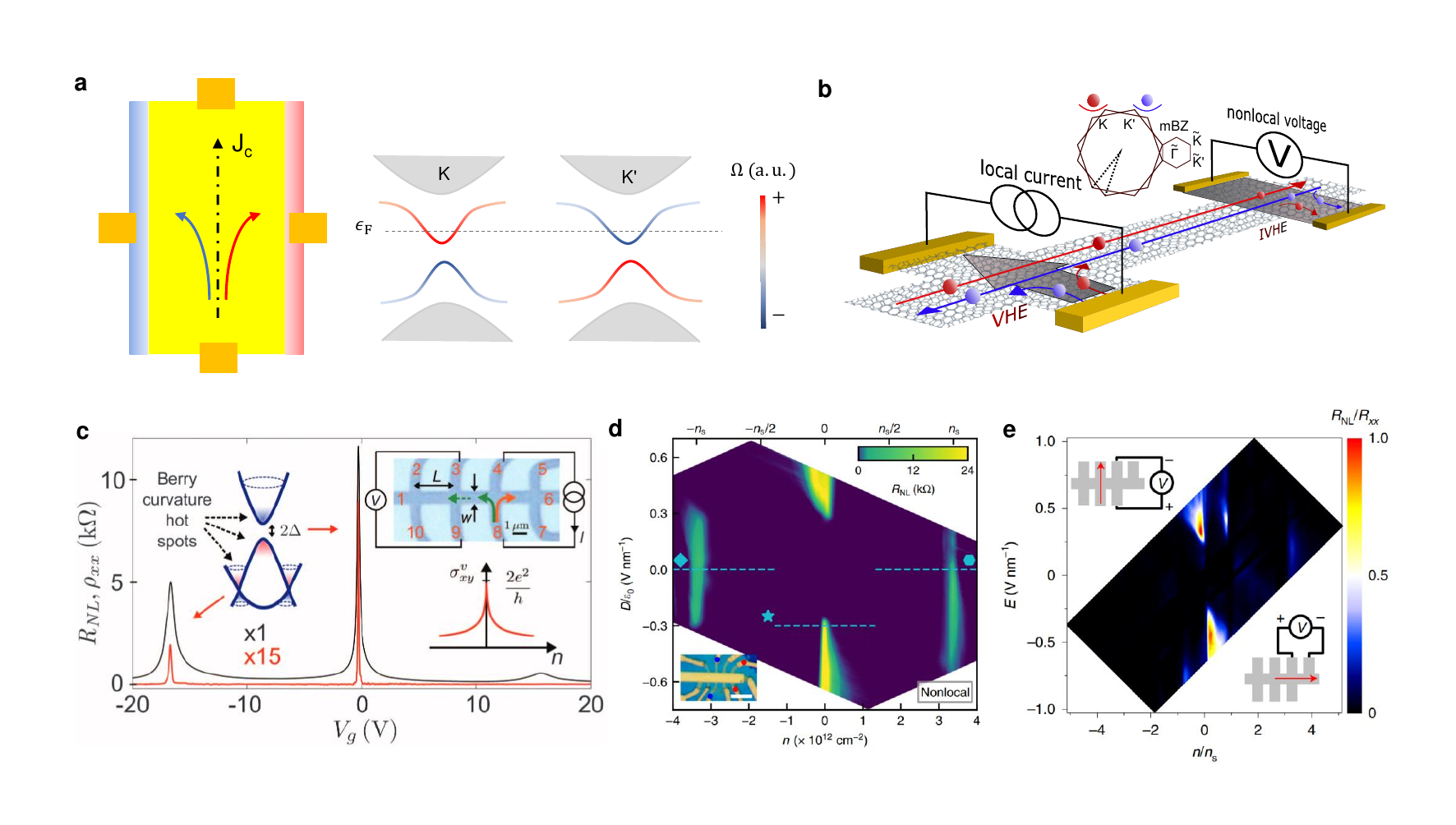}
	\caption[xxx]{\label{fig:valleyhall}{\textbf{ Valley Hall effect probed in electron transport}. 
\textbf{a,} (left) Schematic for the valley Hall effect due to the opposite sign of Berry curvature associated with the two valleys. (right) Panels depict the flat and remote bands with the Fermi energy in the middle of the flat bands. The Fermi energy is tuned with gate electrodes,  resulting in the tuning of the Berry curvature.
\textbf{b,}~The nonlocal resistance measurement scheme to detect the valley Hall effect~\cite{sinha_NC2020_bulk}. 
\textbf{c,} The longitudinal (black) and nonlocal (red) resistance with back gate voltage $V_g$ in Gr/hBN superlattice \cite{gorbachev_S2014_detect}.
The left inset indicates the schematic band structure of the system, with Berry curvature hotspots at the band edges.
\textbf{d,}~Nonlocal resistance as a function of charge density ($n$) and perpendicular electric field ($D/\epsilon_0$) in twisted double bilayer graphene \cite{sinha_NC2020_bulk}. The yellow (blue) color indicates high (low) nonlocal resistance regions.
Insets show the optical image of the device with voltage (current) measuring terminals indicated by red (blue) dots.
The scale bar indicates 5~$\mu$m. 
\textbf{e,} Ratio of nonlocal resistance and local resistance measured in twisted double bilayer graphene \cite{wang_bulk_2022} that suggests a role of edge modes when $R_{NL}\sim R_{L}$. 
\textcolor{black}{Panels \textbf{b} and \textbf{d} are adapted and reproduced, respectively, from Ref.~\cite{sinha_NC2020_bulk}, Springer Nature Ltd. Panel \textbf{c} is reproduced from Ref.~\cite{gorbachev_S2014_detect}, AAAS. Panel \textbf{e} is reproduced from Ref.~\cite{wang_bulk_2022}, Springer Nature Ltd.}
}}
\end{figure}

One of the simplest manifestations of the band structure geometry is the valley Hall effect (VHE) and the associated topological valley current. When inversion symmetry is broken but the time-reversal symmetry is intact, charge carriers can have non-zero Berry curvature with an equal magnitude but opposite signs in two valleys in hexagonal 2D materials.
In accordance with the semiclassical equation of motion ($\bm{\dot{r}}_n=\frac{1}{\hbar}\frac{\partial \varepsilon_n}{\partial \bm{k}}+\frac{e}{\hbar}\bm{E}\times {\bm \Omega}_n(\bm{k})$, see theory box),
electrons originating from the two valleys experience deflections in opposite directions perpendicular to an in-plane electric field (i.e., the source-drain bias field governing the longitudinal charge current).
This sets up a net valley current $J_V=J_K-J_{K'}$, but zero charge current as depicted in Fig.~\ref{fig:valleyhall}a. 
Due to the inverse VHE, the valley current generates a voltage drop across two nonlocal probes (see Fig.~\ref{fig:valleyhall}b) -- thus, the nonlocal resistance $R_\text{NL}$ (nonlocal voltage normalized with the charge current) can be used to probe the valley current.
VHE can be optically probed using Kerr rotation microscopy as well~\cite{lee_electrical_2016}. 

While VHE has been reported using conventional 2D materials with broken-inversion symmetry such as gapped bilayer graphene (BLG)~\cite{shimazaki_NP2015_gen, sui_NP2015_gate, yin2022tunable}, monolayer MoS$_2$, \cite{mak_valley_2014} bilayer MoS$_2$ under electric field, etc., \moire materials provide a versatile platform for Berry curvature hotspots and hence VHE.
Indeed, electrical detection of the bulk valley current was first reported~\cite{gorbachev_S2014_detect} in a \moire system of monolayer graphene aligned to hBN by measuring the nonlocal resistance $R_\text{NL}$ close to the charge neutrality point gap and the two \moire gaps (Fig.~\ref{fig:valleyhall}c).
The advent of twisting 2D materials provides further opportunities to generate and manipulate valley currents. 
As demonstrated in twisted double bilayer graphene, one can have \moire bands with non-zero Berry curvature while also leveraging the perpendicular electric field to tune the bands \cite{adakPRB2020} and hence the valley current \cite{sinha_NC2020_bulk}(see Fig.~\ref{fig:valleyhall}d).

In these experiments, a cubic scaling between the $R_\text{NL}$ and the local resistance $R_\text{L}$ (i.e., longitudinal resistance $R_{xx}$), $R_\text{NL} \propto R_\text{L}^3$ establishes the bulk nature of the valley current~\cite{gorbachev_S2014_detect} .
Careful considerations are required while using non-local geometry to probe bulk valley current. For example, such measurements should be done at elevated temperatures to suppress possible contributions of edge modes. 
At low temperatures, edge modes can contribute \cite{wang_bulk_2022}; 
a notable indication is when $R_{NL}\sim R_{L}$, as seen in Fig.~\ref{fig:valleyhall}e. 

Some experimental studies point toward the signature of higher-order topological "hinge" modes in non-local measurements \cite{ma_moire_2020}, an aspect that needs further confirmation from other probes. 
Furthermore, \moire engineering provides band structures that support stronger Berry curvature hotspots and hence larger valley currents; see Fig.~\ref{fig:comparisonwithmoire} for scaling of Berry curvature with lattice constant. 
The tunability of bandwidth of the \moire flatbands further opens avenues for engineering Berry curvature hotspots \cite{sinha_NC2020_bulk}.

\subsection{Quantum valley Hall effect (QVH) }

\begin{figure}[H]
	\centering
	\includegraphics[width=14cm]{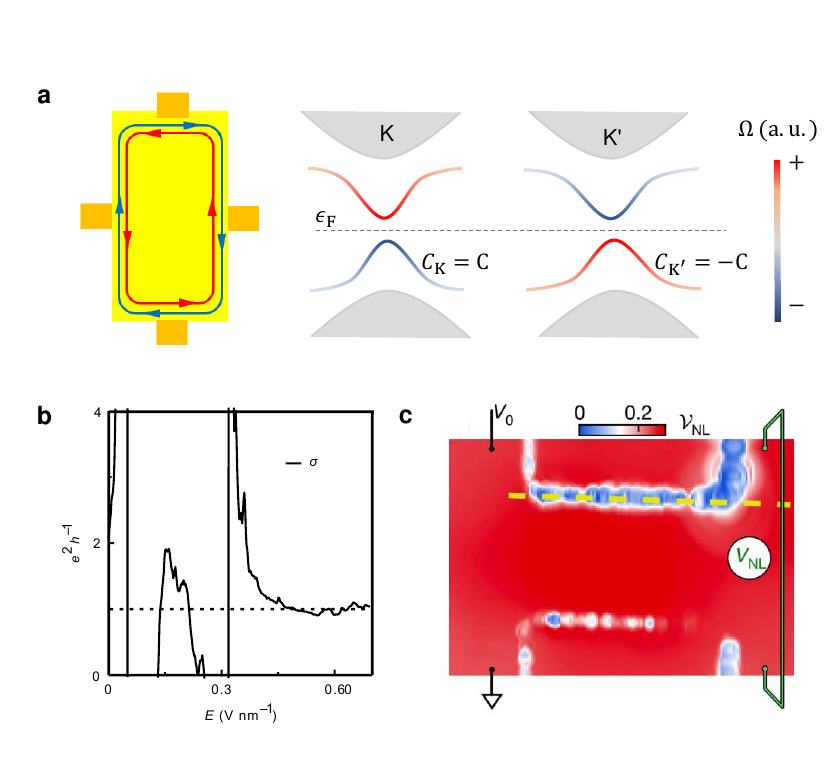}
	\caption[xxx]{\label{fig:quantumvalleyhall}{\textbf{ Quantum valley Hall effect} 
\textbf{a,} (left panel) schematic of two counter-circulating valley Hall edge modes when the valley Chern number is 1, with the total Chern number being zero as the time-reversal symmetry is protected. (right panel) shows the Fermi energy location relative to the flat bands in the presence of time-reversal symmetry. \textbf{b,} \textcolor{black}{Plot of edge conductance ($\sigma$) with perpendicular electric field ($E$), where a diverging $\sigma$ close to $E=0.33$~\Vbynm{} indicates a topological phase transition~\cite{wang_bulk_2022}. In the topological phase, the $\sigma$ is of the order of $e^2/h$}. \textbf{c,} Scanning gate microscopy of a graphene device \cite{aharon-steinberg_long-range_2021} while the current and potential at various contacts are measured. This color scale plot shows the device's non-local voltage and a non-topological edge mode.
\textcolor{black}{Panel \textbf{b} is adapted from Ref.~\cite{wang_bulk_2022}, Springer Nature Ltd. Panel \textbf{c} is reproduced from Ref.~\cite{aharon-steinberg_long-range_2021}, Springer Nature Ltd.}
}}
\end{figure}

\Moire materials provide an interesting opportunity to realize a plethora of quantum valley Hall (QVH) states that directly reveal the topological character of the valley current  ~\cite{ren_topological_2016}.  
QVH states have been predicted in multiple twisted graphene systems~\cite{liu_quantum_2019, zhang_spin-polarized_2022}. 
These systems possess low-energy \moire bands, which may be four-fold degenerate due to valley and spin, and can have non-zero Chern number.
Related by time-reversal, each \moire band with a Chern number $C_K$ from $K$ valley has a counterpart from $K'$ valley with an opposite Chern number $C_{K'} = -C_K$.
Without breaking the valley symmetry, filling two \moire bands from opposite valleys leads to a QVH state characterized by counter-propagating edge states. 
As illustrated in Fig.~\ref{fig:quantumvalleyhall}a, these edge modes travel in opposite directions; the schematic on the right shows the location of the Fermi energy relative to the flat bands. 
Recent studies in twisted double bilayer graphene by combining low-temperature transport and Landauer-Buttiker analysis provide evidence for the QVH effect and topological transition induced by a perpendicular electric field \cite{wang_bulk_2022} (see Fig.~\ref{fig:quantumvalleyhall}b). 

Nonetheless, measuring QVH states via edge transport faces challenges due to valley symmetry breaking from the disruption of the \moire pattern at the edge of the device~\cite{wang_bulk_2022}. 
Combining scanning SQUID and transport measurements on graphene \cite{aharon-steinberg_long-range_2021} provides evidence for long-range edge currents of non-topological origin in 2D systems (see Fig.~\ref{fig:quantumvalleyhall}c). 
Consequently, interpreting studies of topological edge modes in non-local geometries requires careful consideration.

\subsection{Quantum anomalous Hall effect (QAH) \& orbital magnetism}

\begin{figure}[H]
	\centering
	\includegraphics[width=15cm]{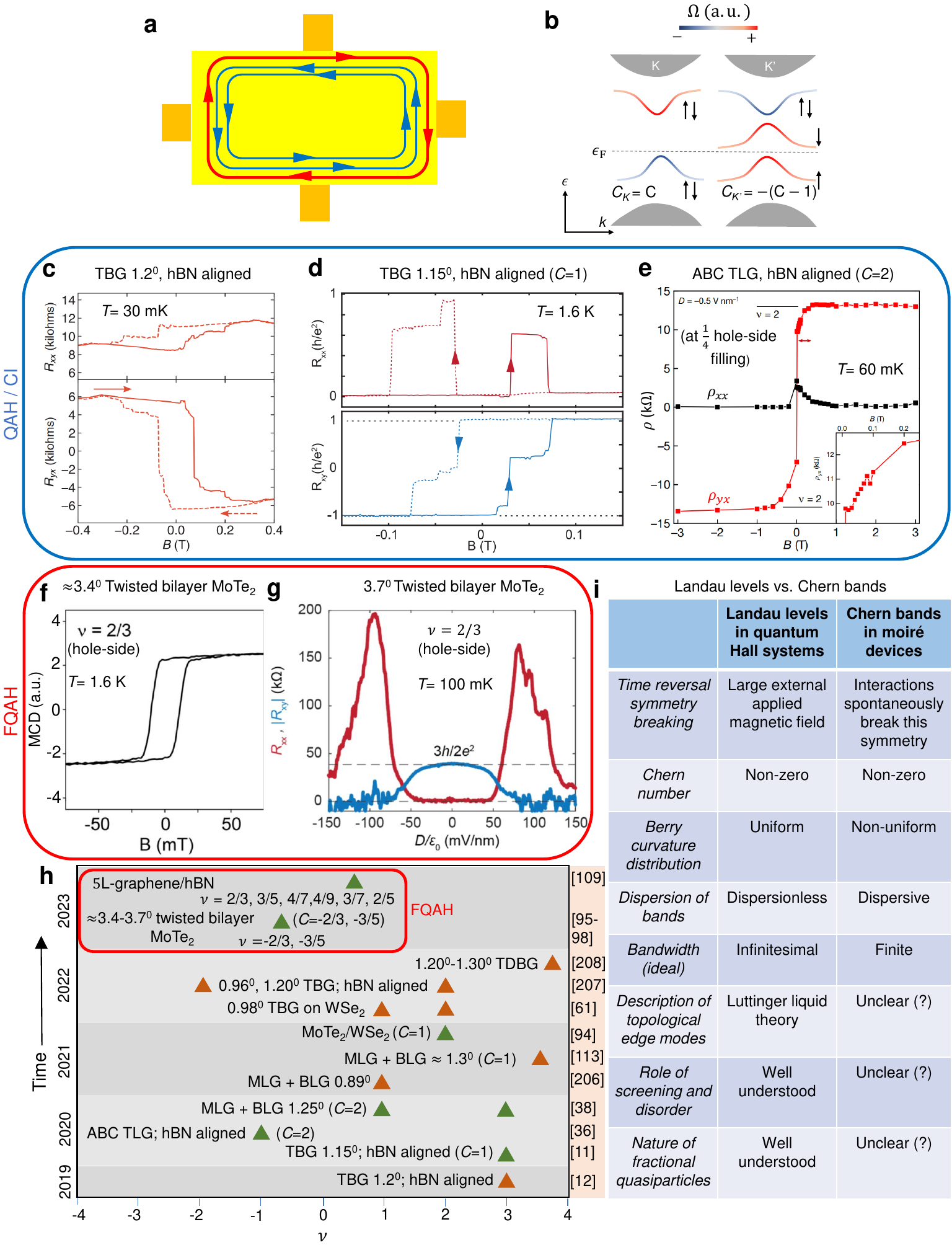}
	\caption[xxx]{\footnotesize \label{fig:QAH}{\textbf{Quantum anomalous Hall effect in \moire superlattices.} 
 \textbf{a,}~Flow of the edge modes while the bulk is insulating. This is analogous to the Quantum Hall state, with key differences, but the QAH/CI state is stabilized at zero magnetic field. 
 \textbf{b,}~Schematic of \moire bands exhibiting quantum anomalous Hall effect. The color indicates the Berry curvature.
 $\epsilon_\textbf{F}$ indicates the Fermi energy.
 \textbf{c, d,}~ Longitudinal resistance \Rxx{} and Hall resistance \Rxy{} in twisted bilayer graphene system from Refs. \cite{sharpe_emergent_2019} and \cite{serlin2020intrinsic} respectively).
 \textbf{e,}~Quantum anomalous Hall effect ($C=2$) in abc-trilayer/hBN superlattice at 1/4 filling ($\nu=-1$) \cite{chen_tunable_2020}.
 The Hall resistivity (red) is quantized concomitantly with zero longitudinal resistivity (black).
 \textbf{f, g,}~Fractional quantum anomalous Hall (FQAH) effect~\cite{cai_signatures_2023, zeng2023thermodynamic,park_observation_2023} in twisted bilayer MoTe$_2$ at a hole-side filling of $v=2/3$.
 The magnetic circular dichroism (MCD) signal in \textbf{f} shows hysteresis with the perpendicular magnetic field (Ref.~\cite{zeng2023thermodynamic}).
 A quantized $R_{xy}=3h/2e^2$ (blue) along with small $R_{xx}$ (red) at 50 mT is observed in \textbf{g} (Ref.~\cite{park_observation_2023}), for a perpendicular electric field $|D|/\epsilon_0 \lesssim 20$~mV/nm.
 \textbf{h,}~A timeline to show various \moire superlattices explored and the corresponding filling factor $\nu$ where quantum anomalous Hall effect and fractional quantum anomalous Hall effect are observed. Green (Brown) triangle indicates quantized (non-quantized) plateaus.
 The Refs. (\cite{sharpe_emergent_2019, serlin2020intrinsic,chen_tunable_2020, polshyn2020electrical, chen2021electrically, polshyn_topological_2022, li2021quantum,lin_spin-orbitdriven_2022, tseng2022anomalous, kuiri2022spontaneous, cai_signatures_2023, zeng2023thermodynamic,park_observation_2023, xu_observation_2023, lu_fractional_2023}) \textcolor{black}{of the corresponding studies are mentioned to the right}.
 \textbf{i,}~A table comparing Landau levels in conventional 2D systems with Chern bands in \moire devices.
 \textcolor{black}{Panels \textbf{c}, \textbf{d} are adapted from Ref.~\cite{sharpe_emergent_2019}, and Ref.~\cite{serlin2020intrinsic}, respectively, AAAS. Panels \textbf{e}, \textbf{f}, \textbf{g} are adapted from Ref.~\cite{chen_tunable_2020}, Ref.~\cite{zeng2023thermodynamic}, and Ref.~\cite{park_observation_2023}, respectively, Springer Nature Ltd.}
    }}
\end{figure}

The quantum anomalous Hall effect (QAH)\cite{liu_quantum_2016, chang_colloquium_2023} is a robust topological phenomenon in which the Hall conductivity ($\sigma_{xy}$) becomes quantized at $\sigma_{xy}=C e^2/h$, without requiring an external magnetic field.  
Here $C$ represents the total Chern number of the occupied bands and quantifies the number of edge modes in the sample (see Fig.~\ref{fig:QAH}a). 
Initially theorized by Duncan Haldane\cite{haldane_model_1988} in 1988, QAH  was first experimentally realized in magnetically doped topological insulators (TI), such as (Bi,Sb)$_2$(Se,Te)$_3$, \cite{chang_experimental_2013} and later in intrinsic magnetic TI, such as MnBi$_2$Te$_4$ \cite{deng_quantum_2020}. 
In both cases, spin-orbit coupling (SOC) played a crucial role.
Due to the notably weak SOC, graphene-based systems are unsuitable for realizing QAH.
Limited success has been obtained by employing the proximity effect with magnetic materials\cite{wang_proximity-induced_2015}.

The recent advent of \moire materials opened new routes to realize the QAH effect.
Here anisotropy of the electron orbitals instead of SOC assumes a pivotal role, leading to orbital magnetism.
First, electron transport measurement by Sharpe et al. \cite{sharpe_emergent_2019} in hBN-aligned twisted bilayer graphene shows pronounced hysteresis in both $R_{xx}$ and $R_{xy}$ around 3/4 filling of the flat band, suggesting an unusual ferromagnetic state in a completely carbon-based material (see Fig.~\ref{fig:QAH}c). Later, a robust zero magnetic field quantized state was observed at 3/4 filling of a TBG aligned to hBN \cite{serlin2020intrinsic} (see Fig.~\ref{fig:QAH}d). Subsequently, the QAH state has been observed in several other twisted graphene systems -- ABC TLG aligned to hBN \cite{chen_tunable_2020} (as seen in Fig.~\ref{fig:QAH}e) and twisted monolayer-bilayer \cite{polshyn2020electrical}.
Fig.~\ref{fig:QAH}h summarizes the \moire systems in which QAH and FQAH have been recently observed at different filling factors $\nu$.

The underlying mechanism for QAH in twisted graphene systems is the availability of Dirac bands that can be readily gapped due to the \moire effect and narrow bandwidth. 
As discussed in the last section, these \moire bands can possess non-zero Chern numbers with opposite signs for two opposite valleys. 
Without breaking the time-reversal symmetry, the Chern number of such a system is zero. The flat bands formed due to the stacking of twisted layers have a narrow bandwidth and high density of states. As a result, there can be situations where the valley symmetry can be spontaneously broken due to the Stoner criterion -- as it is energetically favorable to valley- or spin-polarize the system. The nature of spontaneous symmetry breaking depends on the competition between relevant exchange energy, spin ($J_s$ or valley $J_v$), and the typical spacing in the energy of the orbital states $1/\rho(\epsilon_\text{F})$; here, $\rho(\epsilon_\text{F})$ is the density of states at Fermi energy $\epsilon_\text{F}$. An easy way to think about this is the particle-in-a-box limit of the Stoner criterion, where the system prefers to polarize as the exchange energy leads to an overall reduction in energy even though a higher orbital state is occupied when $J>1/\rho(\epsilon_\text{F})$. 

\Moire structures using hetero and homo bilayers of TMDCs\cite{devakul_quantum_2022}, without needing alignment with hBN, have emerged as a powerful platform for studying topological physics and quantum simulation. 
The impetus for these studies was provided by the demonstrations of \moire based on hetero bilayers of TMDCs as quantum simulators for Hubbard model \cite{wu_hubbard_2018}, with the top \moire valence bands of one TMDC being isolated from the bands of the other TMDC layer. Further studies revealed that homobilayers of certain TMDCs, like MoTe$_2$, lead to a quantum simulator of 2D topological insulators mimicking the Kane-Mele Hamiltonian. TMDC \moire isolated bands have a narrow bandwidth further tunable by a perpendicular electric field. 

Spontaneous symmetry breaking due to a Stoner-like mechanism has led to the observation of QAH effect in  \moire systems such as TMDC heterobilayers \cite{li2021quantum} and homobilayers \cite{cai_signatures_2023, park_observation_2023, zeng2023thermodynamic, xu_observation_2023}. 
The experimental tests for these measurements included standard Hall measurements, compressibility measurements, and magneto-optic measurements to probe the magnetization associated with the QAH effect. The valley-contrasting Chern bands with nontrivial spin texture are formed from interlayer hybridization. While a detailed theoretical understanding of the origin of Chern bands in these \moire systems is still evolving, it is already evident that Chern bands are much more widely available in \moire superlattices than in atomic lattices. Additionally, the non-volatile memory using these Chern insulating states has been demonstrated \cite{polshyn2020electrical} and holds promise for memory devices.

\subsection {Fractional Chern insulator \&  Fractional QAH states}
\textcolor{black}{The fractional quantum Hall (FQH) effect is a remarkable phenomenon where interactions within quantum Hall states result in the quantization of Hall conductivity at fractional values of $e^2/h$. Moiré materials hosting flat Chern bands are emerging as a key platform for observing fractional Chern insulator (FCI) states, which are the lattice analogues of FQH states.}
Several theoretical studies have proposed the emergence of FCI in twisted graphene \cite{ledwith_fractional_2020,repellin_chern_2020, wilhelm_interplay_2021} and twisted TMDCs \cite{wang_fractional_2024, reddy_fractional_2023, reddy_toward_2023, crepel_anomalous_2023, devakul_magic_2021, li_spontaneous_2021}. 
FCI was observed in Bernal-stacked BLG aligned with hBN under a large magnetic field~\cite{spanton_observation_2018}.
Recently, Xie and colleagues have documented the observation of eight FCI states in magic-angle TBG under low magnetic fields~\cite{xie_fractional_2021}. They accomplished this by utilizing precise measurements of local compressibility. 

\textcolor{black}{While FCI states in moiré materials typically necessitate an external magnetic field for stabilization\cite{xie_fractional_2021}, we reserve the term 'fractional quantum anomalous Hall' (FQAH) specifically for fractional states observed without an external magnetic field.}
The flat bands induced by \moire superlattices provide the most promising direction toward the realization of FQAH states \cite{chang_colloquium_2023}. The motivations for this are manifold -- the study of fractional quasiparticles and their topology at zero magnetic field and the topological superconductivity emanating from fractional quasiparticles. 
\textcolor{black}{Rhombohedral pentalayer graphene aligned with hBN~\cite{lu_fractional_2023} has emerged as a host for FQAH states at high displacement fields, where a $C=1$ Chern band hosts the FQAH phases~\cite{dong_theory_2023, zhou_fractional_2023, dong_anomalous_2023}. These new findings are expected to be carried forward to four- or six-layer graphene~\cite{dong_theory_2023, zhou_fractional_2023, dong_anomalous_2023}, and could enable the realization of FQAH and superconductivity in the same device, allowing for novel possibilities.}
\textcolor{black}{Apart from FQAH states, graphene-based \moire{} systems host other exotic states such as charge density wave and symmetry-broken Chern insulator states at fractional filling~\cite{xie_fractional_2021, polshyn_topological_2022, bhowmik_broken-symmetry_2022, he_symmetry-broken_2023}, sustained at zero magnetic field limit.
These states, distinct from the physics of fractionalized quasiparticles, are attributed to translational symmetry breaking of underlying superlattice due to electron-electron interaction.}

The rapid experimental progress in studying TMDC homobilayers of MoTe$_2$ has further led to the observation of the FQAH effect \cite{cai_signatures_2023,park_observation_2023,zeng2023thermodynamic,xu_observation_2023} (see Fig.~\ref{fig:QAH}f,g). 
Experimental techniques involved direct electrical transport \cite{park_observation_2023,xu_observation_2023} with quantization of transverse Hall conductivity (Fig.~\ref{fig:QAH}g), thermodynamic compressibility measurements \cite{zeng2023thermodynamic} and magneto-optic properties (Fig.~\ref{fig:QAH}f)\cite{cai_signatures_2023,zeng2023thermodynamic}. 
Remarkably, these \textcolor{black}{FQAH} states have demonstrated stability up to temperatures as high as 10~K, distinguishing them from \textcolor{black}{Landau level-based fractional quantum Hall} states, which typically have much smaller energy gaps.
This highlights the novel aspects of the \moire platform.

The pioneering studies \cite{cai_signatures_2023,park_observation_2023,zeng2023thermodynamic,xu_observation_2023} of the FQAH effect open up fundamental questions about topology associated with the quasiparticles in FQAH 
and those in fractional quantum Hall studied at high magnetic fields. 
\textcolor{black}{Notably, while 2H-stacked MoTe$_2$ is used in realizing FQAH, Td-stacked MoTe$_2$ superconducts \cite{rhodes_enhanced_2021, jindal_coupled_2023} and has coupled ferroelectric states \cite{jindal_coupled_2023}. Phase engineering demonstrated in other transition metal dichalcogenides could lead to two different phases in close spatial proximity\cite{kappera_phase-engineered_2014}. This opens up the possibility of designing devices with FQAH, superconductor, and ferroelectric phases in close proximity and exploring new physics arising from the interplay of these phases.}
Theoretical proposals for new TMDC systems, like ZrS$_2$, offer the prospect of stabilizing FCI \cite{claassen_ultra-strong_2022} at higher temperatures beyond 10~K. This can usher new applications and opportunities to uncover novel physics. 
\textcolor{black}{Additionally, recent theoretical studies have indicated the possibility of emergent fractional quasiparticles or composite fermions in MoTe$_2$ heterostructures at 1/2 and 3/4 fillings even in the absence of a magnetic field~\cite{PhysRevLett.131.136502,PhysRevLett.131.136501}. However, in contrast to the FQAH states at other fillings, at these even denominator fillings, the composite fermions condense into a gapless Fermi liquid state~\cite{jain_twist_2023}.}

While the topological boundary modes of Chern insulators resemble \textcolor{black}{those arising from Landau levels in a conventional quantum Hall system} -- they both possess a non-zero Chern number -- they are fundamentally very different. 
The Berry curvature distribution in dispersionless Landau levels is uniform, whereas the Chern bands in \moire disperse more and host a non-uniform distribution of Berry curvature. Figure~\ref{fig:QAH}i highlights the key distinction between the Landau levels and the \moire Chern bands. Many aspects, like the effect of disorder, screening, and nature of quasiparticles, are not understood. As we discuss later, these unanswered questions present exciting opportunities for further research. 

\subsection{Nonlinear Hall (NLH) effect} 
 
\begin{figure}[H]
	\centering
	\includegraphics[width=16.5cm]{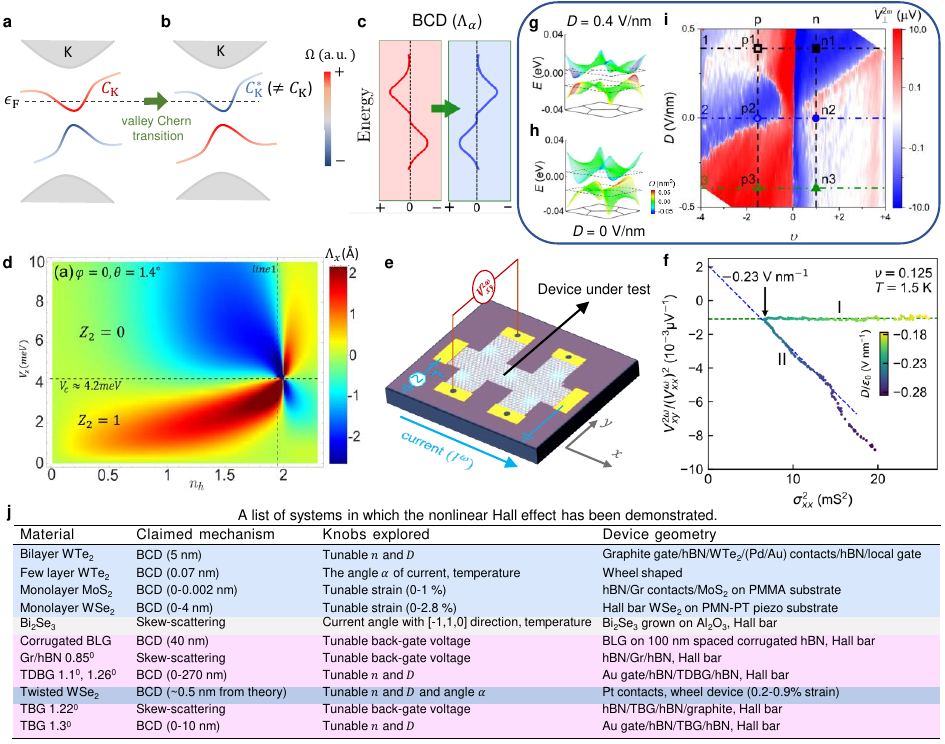}
	\caption[xxx]{\footnotesize \label{fig:NLH}{\textbf{BCD dominated NLH effect in \moire superlattices.}
    \textbf{a, b,}~Schematic representation of a valley Chern transition in the \moire BZ of the K-valley of a \moire superlattice. The transition changes the Berry curvature ($\Omega$) distribution of the flat bands and the valley Chern number of the occupied band from $C_\text{K}$ (\textbf{a}) to a different value $C^{*}_\text{K}$ (\textbf{b}). 
    $\epsilon_{F}$ denotes the Fermi energy. 
    \textbf{c,}~A schematic representation~\cite{chakraborty2022nonlinear} of the impact of a valley Chern transition on the BCD as a function of the chemical potential (or energy). The BCD at a fixed chemical potential (equivalent to fixed filling in experiments) changes signs across the topological phase transition. Breaking of the C$_3$ symmetry (by strain or other means) is essential for supporting a non-zero BCD in the system. It is captured by tilting the bands in a-b (also discussed in Fig.~\ref{fig:knobs}).
    \textbf{d,}~The calculated BCD as a function of the perpendicular displacement field $V_z$ and the occupation number of holes per \moire unit cell $n_h$ in strained twisted bilayer~\cite{hu_nonlinear_2022} WSe$_2$. 
    A change in Z$_2=(C_K-C_{K'})/2$ at $V_z=4.2 meV$ indicates a valley Chern transition.
    For a fixed $n_h$ (dashed line-1), the BCD changes sign as a function of $V_z$.
    \textbf{e,}~ Measurement schematic to probe the second-order NLH voltage \Vxytw. A current $I^\omega$ of low-frequency $\omega$ is sent through a Hall bar device, and the voltage \Vxytw{} is measured perpendicular to the current path. The experiment is done without the application of any external magnetic field.
    \textbf{f,}~The normalized nonlinear Hall voltage \Vxytwnorm{} vs. \sxxsq{}, plotted parametrically with the perpendicular electric field \D{} in $1.1^\circ$ twisted double bilayer graphene at $T=1.5$~K.
    \Vxxw{} is the simultaneously measured longitudinal voltage drop along the current path.
    The change in the sign of \Vxytwnorm{} intercept in the two regimes (I and II) indicates a BCD sign change due to a valley Chern transition~\cite{sinha_NP2022_berry} at $D/\epsilon_0=-0.23$~\Vbynm.
    \textbf{g, h,}~Calculated band structure of twisted bilayer graphene, showing a change in the Berry curvature ($\Omega$) distribution of the \moire bands with perpendicular displacement field $D$.
    \textbf{i,}~Filling ($\nu$) and perpendicular displacement field mapping of NLH voltage $V^{2\omega}_\perp$(=\Vxytw{}), measured in $1.3^\circ$ twisted bilayer graphene at $T=1.5$~K.
    The \Vxytw{} is reported to change due to a change in the Berry curvature distribution of the bands.
    \textbf{j,}~A list of systems (Refs.~\cite{ma_observation_2019, kang_nonlinear_2019, son_strain_2019, qin_strain_2021, he_quantum_2021, ho_hall_2021, he_graphene_2022, sinha_NP2022_berry, huang_giant_2022_published, duan_giant_2022, huang2023intrinsic}) in which the nonlinear Hall effect has been demonstrated. The blue (dark blue) colored columns indicate TMDC (\moire TMDC) based systems. The pink-colored columns indicate graphene-based systems. The white-colored column indicates a topological insulator.
    The knobs explored column indicates different parameters used to vary \Vxytw. Here, $n$ indicates charge density. The device geometry is mentioned from top to bottom along the z-direction.
    \textcolor{black}{Panel \textbf{c} is adapted from Ref.~\cite{chakraborty2022nonlinear}, IOP. Panel \textbf{d} is adapted from Ref.~\cite{hu_nonlinear_2022}, Springer Nature Ltd. Panel \textbf{e-f} is adapted from Ref.~\cite{sinha_NP2022_berry}, Springer Nature Ltd. Panels \textbf{g}, \textbf{h} (\textbf{i}) are adapted (reproduced) from Ref.~\cite{huang2023intrinsic}, APS.}
    }}
\end{figure}

So far, we have focused on linear charge Hall responses, which vanish in a system that preserves time-reversal symmetry. 
However, if such a nonmagnetic system breaks the inversion symmetry, the second-order nonlinear Hall (NLH) effect becomes the dominant response.
As the theory box highlights, the NLH response is primarily induced by BCD, with additional contributions from disorder effects.  
The NLH effect has been recently used (see Fig.~\ref{fig:NLH}) to probe band topology~\cite{ma_topology_2021} and topological phase transitions~\cite{sinha_NP2022_berry, chakraborty2022nonlinear}. 
Its thermoelectric and thermal counterparts have also been predicted to probe Berry curvature distribution and band topology in the system~\cite{du2021nonlinear, chakraborty2022nonlinear}.  

The second-order nonlinear Hall current density is closely connected to the first moment of Berry curvature, namely, the Berry curvature dipole (BCD).
The connection was highlighted  theoretically~\cite{sodemann_PRL2015_quant} by Sodemann and Fu.
For 2D materials in the $x$-$y$ plane, only the out-of-plane component of the Berry curvature (${\bf \Omega}=\Omega_z\hat{z}$) is finite. 
For such systems, the nonlinear current density can be vectorially written as~\cite{sodemann_PRL2015_quant} 
${\bm j} \propto \uvec{z} \times {\bm E}~({\bm \Lambda} \cdot 
{\bm E})$.
Here, the BCD is a pseudo vector with components ($\Lambda_{a}$). For a particular band, it is given by, $\Lambda_{a}=\int \dfrac{d{\kbm}}{(2\pi)^2}\frac{\partial \Omega_z}{\partial k_{a}}f(\epsilon_{\kbm})$, where $a$ denotes a spatial coordinate ($x$ or $y$), and $f(\epsilon_{\kbm})$ is the Fermi-Dirac distribution function. 
Physically, finite $\Lambda_{a}$ arises from the non-uniform Berry curvature distribution over the Fermi surface. 
A finite nonlinear $\bm{j}$, due to BCD, generates a nonlinear Hall voltage \Vxytw{} measured via the lock-in technique, with twice the frequency ($\omega$) of the channel current (Fig.~\ref{fig:NLH}e). 

In the table of Fig.~\ref{fig:NLH}j, we provide a list of systems where the nonlinear Hall voltage \Vxytw{} has been measured and tuned.
In TMDC-based~\cite{ma_observation_2019, kang_nonlinear_2019} 2D materials such as WTe$_2$, the BCD originates due to band tilting, causing a non-uniform Berry curvature distribution.
Electrical knobs such as charge density $n$ and perpendicular electric displacement field $D$ were used to tune the NLH voltage in bilayer~\cite{ma_observation_2019} WTe$_2$ and confirm the BCD-dominated mechanism.
Strain tuning of BCD was demonstrated in MoS$_2$ and WSe$_2$ by straining the substrates~\cite{son_strain_2019, qin_strain_2021}.

A finite BCD ($\Lambda$) in graphene and TMDs-based \moire systems~\cite{sinha_NP2022_berry, huang_giant_2022_published, huang2023intrinsic} arises due to strain in the samples~\cite{kazmierczak_strain_2021, turkel2022orderly} that breaks the C$_3$ rotation symmetry. 
As schematically shown in Fig.~\ref{fig:NLH}a, a non-zero strain engineers the \moire bands to create a non-zero BCD. 
The BCD is proportional to the tilt of the Dirac cones, which in turn is proportional to strain~\cite{sodemann_PRL2015_quant}.
Recent calculations suggest a large BCD in strained twisted bilayer graphene~\cite{pantaleon_tunable_2021, zhang_giant_2022}, and other TMDC based superlattices~\cite{he_giant_2021, hu_nonlinear_2022}, where~\cite{mannai_twistronics_2021}
$
{\rm BCD} \propto v_t  \propto l^{3}(\epsilon_{xy}). 
$
Here, $v_t$ is a tilt parameter in the Dirac Hamiltonian (see theory box), $l$ is the \moire wavelength, and $\epsilon_{xy}$ is a strain parameter.
This simple estimate indicates that the BCD is larger for systems with i) a large moir\'e wavelength (consistent with dimensional analysis; see Fig.~\ref{fig:comparisonwithmoire}) and ii) a large strain.

As shown schematically in Fig.~\ref{fig:NLH}a, b, the bands of \moire superlattices can have non-zero valley Chern number, further tunable by a vertical displacement field~\cite{zhang_nearly_2019}. 
Importantly, the BCD changes sign across the valley Chern transition~\cite{hu_nonlinear_2022, chakraborty2022nonlinear} (schematics of Fig.~\ref{fig:NLH}c), enabling a new tool to probe such transitions using the NLH response without explicitly breaking the time-reversal symmetry.
A change in Z$_2=(C_K-C_{K'})/2$ index characterizes a valley Chern transition.
For example, the calculated BCD (Fig.~\ref{fig:NLH}d) for a strained twisted bilayer WSe$_2$ shows a sign change~\cite{hu_nonlinear_2022} with perpendicular electric field across a valley Chern transition of Z$_2=1$ to Z$_2=0$.
A valley Chern transition of Z$_2=2$ to Z$_2=0$ with the perpendicular electric field has also been demonstrated in twisted double bilayer graphene~\cite{sinha_NP2022_berry}, showing up as a change in sign of the \Vxytwnorm{} intercept (Fig.~\ref{fig:NLH}f).
Such a BCD sign change is also theoretically predicted in BiTeI when it transitions from a trivial to a topological insulating phase~\cite{facio2018strongly} as pressure is tuned.

The nonlinear Hall voltage has recently been measured in other \moire systems such as twisted bilayer graphene, graphene/hBN superlattices, twisted WSe$_2$, etc. (see Fig.~\ref{fig:NLH}j).
Some experiments suggest that both scattering~\cite{he_graphene_2022, duan_giant_2022} and BCD~\cite{sinha_NP2022_berry,huang2023intrinsic, huang_giant_2022_published} mechanisms play an important role in the NLH response of \moire superlattices. 
In twisted bilayer graphene, the redistribution of the Berry curvature hotspot in the \moire bands with perpendicular electric field tunes the BCD~\cite{huang2023intrinsic} dominated \Vxytw{} (Fig.~\ref{fig:NLH}g-i).
Another study~\cite{duan_giant_2022} points to the disorder-induced skew-scattering mechanism as the most prominent close to the \moire superlattice gaps.
Recent developments in the field suggest probing a simplified scaling relation of the form $\frac{E_{xy}^{2\omega}}{(E_{xx}^{\omega})^2}=\frac{\sigma_{yxx}^{(2)}}{\sigma_{xx}}=A\sigma_{xx}^2+B$, to distinguish the various intrinsic and extrinsic contributions to the NLH response.
Here \Exytw{} ($=\frac{V_{xy}^{2\omega}}{width}$) is the generated nonlinear Hall electric field, \Exxw{} ($=\frac{V_{xx}^{\omega}}{length}$) is the longitudinal electric field, and $\sigma_{xx}$ is the longitudinal conductivity.
This scaling relation was probed using the perpendicular electric field as a parameter in Fig.~\ref{fig:NLH}f and temperature as a parameter in few-layer~\cite{kang_nonlinear_2019} WTe$_2$ to extract the BCD. 
The different contributions to the second-order conductivity $\sigma_{yxx}^{(2)}$ have been discussed in Refs.~\cite{du_NC2019_dis, lahiri2022nonlinear}.
More comprehensive experiments exploring the different knobs available in \moire materials might add more insights into different mechanisms generating nonlinear Hall response. 

Recent experiments with TDBG~\cite{sinha_NP2022_berry}, WTe$_2$/WSe$_2$ \moire heterostructures~\cite{kang_switchable_2023}, and other non-\moire materials such as few-layer~\cite{xiao2020berry} WTe$_2$, topological insulator~\cite{nishijima2023ferroic} Pb$_{1-x}$Sn$_x$Te, demonstrate a ferroelectric-like switching of the nonlinear Hall voltage with a perpendicular electric field. 
This observation of hysteresis points to a coupling between band topology and ferroelectric orders and encourages 
exploration of Berry curvature-based next-generation memory devices.

\subsection{Chiral Berry plasmon \& magnetic field free Faraday effect}

In addition to the novel transport phenomena, valley contrasting Berry curvature in materials can also give rise to interesting optical properties, including the emergence of Chiral Berry plasmons (CBPs)~\cite{song_chiral_2016, kumar_chiral_2016}. Plasmons, the collective density oscillations of electrons in a solid or plasma, are valuable for manipulating light at subwavelength scales. Chiral plasmons offer the advantage of nonreciprocity, making them very useful for novel optical instruments such as optical isolators and circulators. 
Typically, chirality in a material is accomplished through symmetry breaking induced by an external magnetic field. However, the uncompensated Berry curvature flux in ferromagnetic materials can generate 
chiral plasmons at the boundaries \cite{song_chiral_2016} even without a magnetic field. Alternately, uncompensated Berry curvature in optically pumped, using circularly polarized light, gapped Dirac materials with valley contrasting Berry curvature can also induce chirality in the edge plasmon modes \cite{song_chiral_2016, kumar_chiral_2016}, enabling on-chip applications. These CBP  waves traversing in opposite directions at the system boundaries have split energy dispersion, with the difference being proportional to the uncompensated Berry curvature flux in the system.

Recent experiments have successfully demonstrated CBPs in twisted bilayer graphene (TBG) with optical pumping alongside other slow bulk plasmon modes. Researchers employed a 1.08-degree TBG to probe the slow plasmon modes in \moire systems and demonstrated a low-energy CBP mode induced by a circularly polarized light~\cite{huang_observation_2022}. 
A finite valley polarization was confirmed by demonstrating a large magnetic field-free Faraday effect.

\subsection{Quantum metric induced non-reciprocal plasmons and  phase stiffness in flat band superconductors}

Furthermore, \moire heterostructures also support extremely tunable intra- and inter-band plasmons\cite{PhysRevB.106.155422}, as investigated in recent experiments\cite{Niels_Nphys21,huang_observation_2022}. 
In a related development, bulk non-reciprocal plasmons in graphene induced by a drift current, mimicking Fizeau drag, with a different propagation wavelength in opposite directions, have been recently demonstrated \cite{Zhao2021}. Spontaneous time-reversal symmetry breaking by interaction effects in \moire heterostructures can also induce intrinsic non-reciprocal plasmons without the Faizeu drag of drifting electrons.  \cite{arora_quantum_2022,  PhysRevLett.125.066801, PhysRevB.106.155422, Dutta_intrinsic_2023}. \textcolor{black}{The quantum metric, referred to as the Fubini study metric, plays a vital role in inducing this non-reciprocity.}

{ \textcolor{black}{The quantum metric ${\cal G}$ is defined by the anti-commutator of the spatial coordinates\cite{ bhalla_PRL2022_reso}, ${\cal G}_{ab}  = \{r_a, r_b\}/2$, which are expressed in terms of the Berry connection as $r_a = {\mathcal R}^a_{pm}$ and $r_b = {\mathcal R}^b_{mp}$ (see theory box for more details). Physically, the Berry curvature captures the band topology, and the quantum metric captures the geometry of the eigenstates. The Berry curvature and the quantum metric are also related to the distance between two quantum states differing by a small momentum, $ds^2 = {\cal Q}_{\mu \nu} dk_\mu dk_\nu$, where ${\cal Q}_{\mu \nu} = \langle \partial_\mu \psi| \partial_\nu \psi \rangle  - \langle\partial_\mu \psi| \psi \rangle \langle \psi| \partial_\nu \psi \rangle$ is the quantum geometric tensor. The real (symmetric) part of the geometric tensor is the quantum metric, and the imaginary  (anti-symmetric) part is the Berry curvature \cite{rossi_quantum_2021}. Both appear in physical quantities, capturing the impact of interband coherence in quantum materials or quantities involving the band overlap function.}

\textcolor{black}{For example, treating light-matter interaction perturbatively within the length gauge typically leads to products of the Berry connection, which appear in several transport and optical phenomena in the form of Berry curvature, the quantum metric, and other higher-rank band geometric tensors\cite{ bhalla_PRL2022_reso,mandal2023intrinsic}. While we have already discussed several examples of Berry curvature-induced phenomena, the quantum metric has been shown to dictate the nonlinear Hall effect in bipartite antiferromagnets \cite{gao_quantum_2023,wang_quantum_2023}. This contribution was recently used to explore the nonlinear valley Hall effect in systems with inversion symmetry \cite{das2023nonlinear}. A different quantum metric-induced contribution captures the effect of interband coherence on the longitudinal nonlinear transport \cite{lahiri_arxiv2023_intrinsic}. In the optical regime, the quantum metric plays an important role in second harmonic generation \cite{bhalla_PRL2022_reso}. More recently,  quantum metric has also been shown to contribute to linear optical conductivity, which manifests as intrinsic contributions in quantum capacitance and dielectric constant\cite{komissarov2023quantum}. Beyond impacting the current, the quantum metric also influences the fluctuations in the current and contributes to thermal and shot  noise\cite{PhysRevLett.130.036202}.}

More interestingly, in any symmetry-breaking collective ground state, which breaks the $U(1)$ symmetry, the quantum metric can be connected to the `phase stiffness' of the collective ground state \cite{rossi_quantum_2021}. Mathematically, this phase stiffness is encoded in the current-current correlation of the system, which captures the system's response to the effective gauge field or the effective vector potential. This was first demonstrated for superconductors \cite{Peotta2015, PhysRevLett.117.045303, PhysRevB.95.024515, Torma_NPhys_Review_22}, and has been experimentally probed in \moire superconductors \cite{Tian_Nature_23}. The finite quantum metric induced superconducting phase stiffness is crucial for supporting superconductivity in flat band \moire heterostructures, and it also dictates the superconducting transition temperature \cite{Torma_NPhys_Review_22, PhysRevX.9.031049, pnas_Mao_23}. }  

\subsection{Nonlinear optical response}

Besides their exciting transport properties, \moire materials offer an attractive platform for engineering nonlinear optical and optoelectronic responses \cite{arora_strain-induced_2021}. 
These are facilitated by the strong light-matter interaction, the presence of a large density of states, and electrically tunable bandgap $\sim$~10~meV supporting photo-response in the sought-after infrared and terahertz regime (see Ref.~\cite{science.adg0014} for a recent review). 
For example, a significant infrared bulk photovoltaic effect in twisted double bilayer graphene (around $\sim$ 3.7 V/W at 7.7 $\mu$ m) was recently demonstrated\cite{ma_intelligent_2022}. They found that the magnitude of the bulk photovoltaic effect was sensitive to the gate voltages along with the polarization and amplitude of light. More interestingly, they used machine learning algorithm-based analysis to use observed photocurrent to identify the `fingerprint' of the incident light in terms of its amplitude and polarization. 

In addition to the bulk photovoltaic effect, hexagonal \moire materials also offer the interesting possibility of selectively exciting the carriers of one valley via circularly polarized light\cite{mak_NN2012_control,huang_observation_2022} to probe large valley polarized anomalous Hall effect and spin Hall effect in spin-valley locked \moire heterostructures. Another interesting impact of the \moire potential in transition metal dichalcogenide heterobilayer (MoTe$_2$/WTe$_2$) is that it can induce a sizable periodic pseudo-magnetic field on the valence band, which induces a large valley contrasting Berry curvature flux in the two valleys. \cite{hu_berry_2023}. The large Berry curvature combined with the spin-valley locking can induce the spin Hall effect, which has been recently observed in experiments \cite{tschirhart_intrinsic_2022, Giant_SHE_23}. Interestingly, the pseudo magnetic field and the valley-contrasting Berry curvature distribution can be probed through the split in dual peaks of the shift currents (dc currents induced by linearly polarized light) as a function of frequency\cite{hu_berry_2023}. The shift current in \moire materials is induced by a third rank band geometric quantity (called the metric/symplectic connection) which is large in \moire materials \cite{bhalla_PRL2022_reso, PhysRevResearch.4.013209} and sensitive to interaction-induced renormalization of the \moire bandstructure \cite{PhysRevResearch.4.013164}.

\section{Challenges in unraveling Berry physics in \moire materials}

\begin{figure}[H]
	\centering
	\includegraphics[width=16cm]{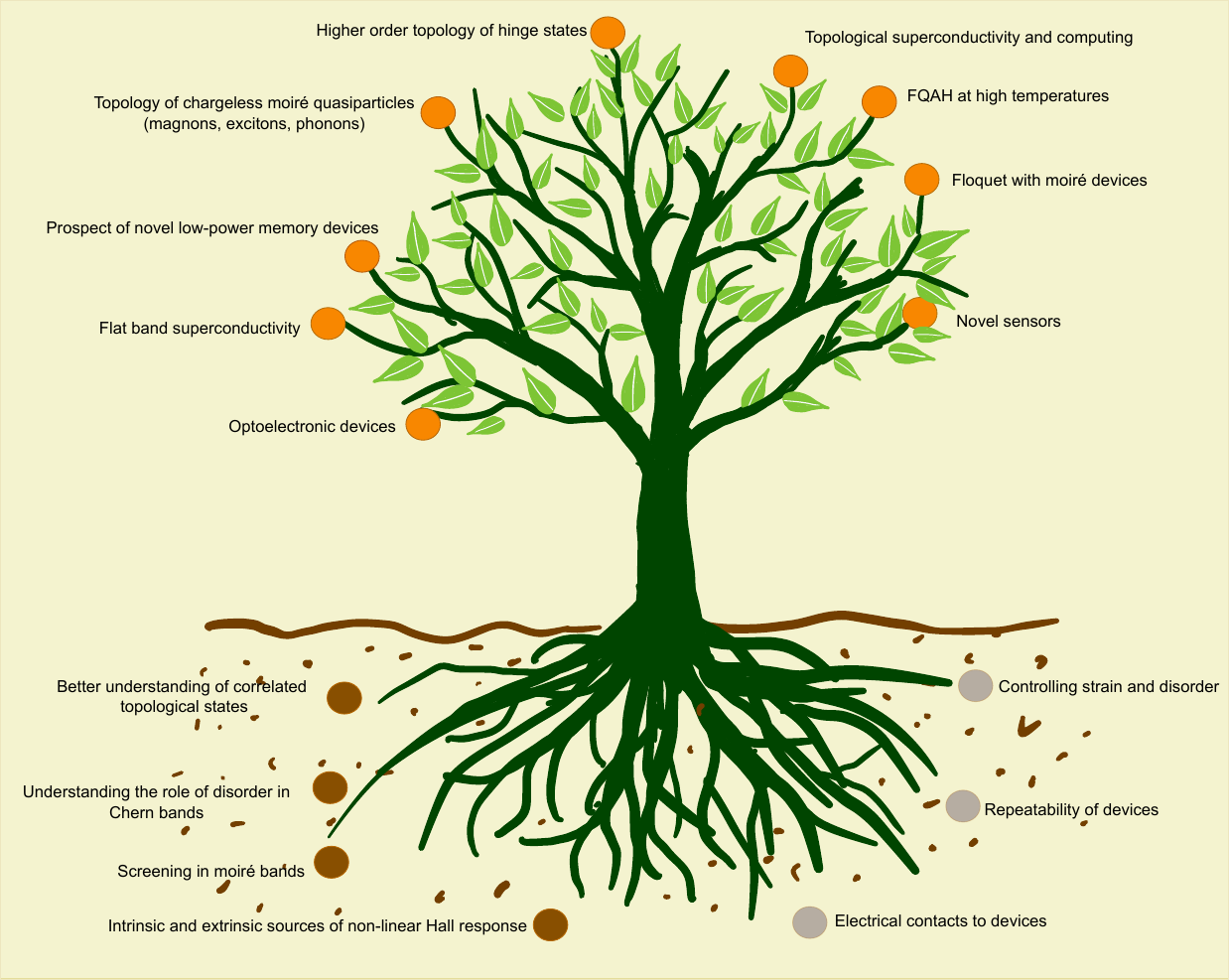}
	\caption[xxx]{\label{fig:summary}
 {{\bf Opportunities and outstanding challenges in \moire materials}. The fruits depict opportunities. The points near the roots indicate current challenges at the device level (grey) and in theoretical understanding (brown).}
}
\end{figure}

As discussed, \moire systems have significantly advanced our understanding of topological phenomena. 
However, the field faces challenges on two critical fronts: device-level obstacles that must be overcome to fully exploit the field's potential and conceptual hurdles that require further advancement.  
Figure~\ref{fig:summary} shows these challenges as aspects under the ground level, close to the roots of the "tree"  that represents this field of the topological \moire  -- improvements in these aspects will strengthen the foundation of this promising field. 

\subsection{Challenges at device level}

One critical challenge revolves around the nature of electrical contact with moiré devices when probing the electrically measurable consequences of Berry curvature.
The contact with these devices can be non-Ohmic at times, resulting in the measurement of various harmonics during modulated measurements using the lock-in technique, even in the absence of Berry curvature-induced effects.
The non-Ohmic contact issue becomes particularly critical in non-linear Hall measurements, where higher harmonics are employed to detect the Berry curvature dipole effect. Heterostrain within \moire devices can lead to spatial inhomogeneity, and this impedes the observation of Berry curvature effects as the band structure of the system will vary over space. It is important to note that spatially non-uniform strain configuration can pose challenges, whereas spatially uniform strain can be a very powerful knob to study Berry physics. Uniform strain can break symmetries and consequently lead to observations that are not seen in a strain-free system. Aspects of \moire physics get affected by any modification of the structure of the twisted layers away from idealized rigid 2D sheets -- resulting in a change in the band structure and the resulting Berry curvature distribution. As a result, reconstruction of the twisted layers and relaxation of the twist will adversely affect the study of Berry physics in the \moire devices. In general, device reproducibility \cite{lau_reproducibility_2022} remains a major obstacle within this field, largely stemming from fabrication challenges.

\subsection{Conceptual and theoretical challenges} 
Besides the experimental challenges, our conceptual understanding and modeling of several facets of \moire physics need more work for further development.
\textcolor{black}{ A comprehensive understanding of distinct symmetry broken phases and the interplay between them in different \moire heterostructures is still lacking. For example, twisted bilayer graphene exhibits strongly correlated states and superconductivity, but the interplay between these two phases is poorly understood. The origin of ferroelectricity in \moire heterostructures is unclear, and their interplay with band geometry needs to be probed further. Understanding the role of the distribution of quantum geometry in momentum and energy space in inducing and controlling strongly correlated and FCI phases needs to be probed further.}

\textcolor{black}{In the context of non-linear transport studies in \moire systems and 2D materials in general, the contributions of various scattering mechanisms like skew scattering and side jump together with their dependence on temperature in flat band systems are not fully understood; aspects important to understand the intrinsic and extrinsic contributions to non-linear Hall response. How these extrinsic mechanisms of non-linear Hall response depend sensitively on the electric field tunable bandstructure of \moire materials is poorly understood thus far.} 

\textcolor{black}{A few specific challenges related to the recently discovered FCI states are as follows. Firstly,  our understanding of the nature of boundary modes at the edge of twisted layers and their robustness is incomplete. 
There are indications of valley mixing \cite{khalifa_absence_2023}, which could affect connections with experimental observations. For the quantum Hall states, the Luttinger liquid theory has been very successful, and it is unclear whether it can capture the physics of topological boundary modes in CI and FCI. Secondly, the consequence of disorder on Chern bands is not fully understood\cite{parameswaran_fractional_2013}. 
Landau levels do not disperse, unlike Chern bands that have finite bandwidth. In addition, the distribution of Berry curvature is uniform for Landau levels, whereas it has structures in momentum space for Chern insulators \cite{liu_recent_2023}. Mixing of Landau levels has a bearing on the overall phase diagram of fractional quantum Hall systems, and details of mixing in \moire Chern bands are likely to be fundamentally different because of the difference in dispersion and distribution of Berry curvature. Lastly, the microscopic nature of screening and the range of interactions in Chern bands have implications for the phase diagram of CI and FCI. The competing phases in \moire are affected by the microscopic details of screening, which, in turn, has implications for studies of FCI.}

\textcolor{black}{Having considered some of the challenges in the studies of band geometry in \moire systems, we next point out their exciting possibilities and prospects.}

\section{Exciting possibilities}

\Moire heterostructures have emerged as an extremely tunable, versatile, and robust platform for quantum simulators by enabling tunability, control, and engineering of topology, strong correlations, and band structure \cite{kennes_moire_2021}. Stacking and twisting different 2D materials gives rise to various ordered phases and associated low-energy Hamiltonians. These include the Mott insulator, Kondo insulator, magnetism, superconductivity, charge and spin density waves, superconductivity, orbital magnetism, interfacial ferroelectricity,   and topological flat bands. This opens up a  new frontier for studying these quantum phenomena in a controlled manner. A distinct advantage of \moire materials is their extreme tunability and more manageable temperature scales, overcoming the limitations of conventional crystalline solids and ultracold atoms as quantum simulators. Figure~\ref{fig:summary} shows the "fruits" of the "tree"  that represent promising prospects for the field of  \moire for exploration of Berry physics and topology --  we discuss these next.

\subsection{Higher-order topology}

In addition to the valley Chern topology, \moire materials are promising candidates for hosting higher-order topology and the associated corner modes. For example, a robust second-order topological insulating state has been predicted in small angle and large commensurate angle TBG and twisted hBN\cite{park_higher-order_2019,liu_higher-order_2021}, along with other  bilayers~\cite{liu_second-order_2022}. 
These proposals highlight the importance of higher energy bands (and bandgaps) in \moire materials, with the higher order insulator state originating either in the gaps at integer fillings or at gaps opening at half filling. 
It is believed that a statistical ensemble of such zero-dimensional corner modes can weakly couple to form delocalized and diffusive quasi-one-dimensional channels, offering an experimental signature in terms of non-quantized nonlocal resistivity \cite{ma_moire_2020}. 
This adds higher-order topology to the list of quantum phenomena that can be simulated in \moire materials. 

\subsection{Floquet-engineered Berry curvature and Floquet-topological phase}
 
Floquet engineering offers a route to ultrafast switching of topological and other phases and additional knobs for experimental tuning of interesting physical properties \cite{rodriguez-vega_low-frequency_2021}. However, incompatible time and energy scales limit its impact on conventional materials. \Moire system overcomes this limitation by providing opportunities to electrically tune the band gap, enabling Floquet engineering of nontrivial topology and other possibilities. For example, there are predictions that twisted bilayer graphene near the magic angle driven by circularly polarized laser pulses can support optically controllable topological Floquet band structure having Berry curvature analogous to a valley Chern insulator ~\cite{topp_topological_2019, li_floquet-engineered_2020}. Similar light-induced valley-polarized flat Floquet Chern bands and tunable large Chern numbers have also been predicted in twisted multilayer graphene\cite{lu_valley-selective_2021, rodriguez-vega_low-frequency_2021}, and twisted transition metal dichalcogenide bilayers\cite{vogl_floquet_2021, rodriguez-vega_low-frequency_2021}. These Floquet-engineered topological transitions are enabled by tuning interlayer coupling strength in \moire heterostructures, which the optical field's amplitude, polarization, and frequency can experimentally control.

\subsection{Other topological chargeless excitations: \Moire Excitons, phonons and magnons} 

The tunable periodic \moire potential in \moire hetero-bilayers can significantly alter the exciton bands and optical spectrum and pave the way for optoelectronic properties \cite{wu_topological_2017, brem_nanoLett_20, Shen_Exciton_NNat23} tailored for specific needs. 
In the presence of time-reversal symmetry breaking induced by a Zeeman field, \moire hetero-bilayers can simultaneously support twist angle-dependent satellite exciton peaks and topological exciton bands\cite{wu_topological_2017}. 
\Moire materials also serve as a unique playground for exploring bosonic topology via their chiral phonons\cite{nischay_Nanolett_2021,Liu_nanoLett_2022,Lin_acsnano_18, Li_review_2023} and magnon bands\cite{li_moire_2020,Ganguli_nanolett_23}, leading to distinct thermal transport signatures. 
For example, the \moire magnon bands in twisted antiferromagnetic bilayers coupled ferromagnetically have been predicted to be flat and topological for a wide range of twist angles \cite{li_moire_2020}. 
The Berry curvature associated with the magnon bands provides further opportunities for transverse thermal spin transport.

In addition to the possibilities mentioned above, \moire materials also offer exciting prospects to probe topological charge pumping \cite{fujimoto_topological_2020, zhang_topological_2020, su_topological_2020}, real space Berry curvature\cite{zhai_theory_2020} effects, spin Hall insulators\cite{wu_topological_2019_2, claassen_ultra-strong_2022}, and topological superconductivity  \cite{classen_competing_2019,wu_topological_2019}.  

\subsection{Novel \Moire devices} 
Currently, these technology demonstrations work under less-than-optimal conditions for room-temperature applications.
Nevertheless, leveraging their unique topological nature, \moire devices present intriguing possibilities.
For example, spin torque magnetic memories, which utilize electrically actuated spin currents to manipulate magnetic moments, could benefit from the spin Hall torque demonstrated near $\nu = 1$ and $\nu = 2$ in MoTe$_2$/WSe$_2$ heterostructures stemming from their large Berry curvature\cite{tschirhart_intrinsic_2022}.
Record low current densities required for switching demonstrated using this platform could enable ultra-low-power switching. 
Combining light with the Chern insulators has also been proposed as a platform for optical memory elements \cite{pershoguba_optical_2022}. 
The tunable bands offer opportunities for sensing. 
Recent research has revealed that twisted double-bilayer graphene exhibits a substantial photoresponse sensitive to polarization. 
Combining this large photoresponse and machine learning principles, an IR detector quantifying intensity and polarization has been demonstrated \cite{ma_intelligent_2022}. Principles of geometric deep optical sensing\cite{yuan_geometric_2023} have recently garnered significant attention, and leveraging the quantum geometry of \moire devices holds great potential for further advancements in sensing technology. 
Additionally, the non-linear Hall response of \moire materials offers opportunities for broadband energy harvesting, extending the proposals for 2D crystals \cite{onishi_high-efficiency_2023}. Due to its broadband nature, quantum rectification based on nonlinear Hall response may become a crucial technology for harnessing RF, millimeter, and THz waves for wireless charging and energy harvesting. Besides the possibilities discussed here, topological \moire materials can also contribute to the ongoing development of quantum technologies \cite{montblanch_layered_2023}.

This review presents a pedagogical introduction to \moire materials, highlighting their versatility as unique platforms for exploring Berry physics.  We have elucidated key experiments and fundamental ideas central to topological moiré physics. Additionally, we have highlighted the challenges to the development of this field and the exciting prospects and opportunities it offers for fundamental science as well as practical device applications.

%-------------------------------------

\section*{Highlighted references}
1. \textbf{This review highlights leveraging the tunability of \moire heterostructures to experimentally simulate different fundamental many-body quantum models in condensed matter.}\\
Kennes, D.M. et al.~\Moire{} heterostructures as a condensed-matter quantum simulator~\cite{kennes_moire_2021}.\\
2. \textbf{A theoretical study identifying possible 2D materials.}\\
Mounet, N. et al. Two-dimensional materials from high-throughput computational exfoliation of experimentally
known compounds~\cite{mounet_two-dimensional_2018}.\\
3. \textbf{The review provides an overview on the emerging 2D materials research and outlines the future directions.}\\
Geim, A. K., Grigorieva, I. V. Van der Waals heterostuctures~\cite{geim_van_2013},\\
Novoselov, K.S., Mishchenko, A., Carvalho, A., Neto, A. H. C. 2D materials and van der Waals heterostructures~\cite{novoselv2018vdWheterostructures}.\\
4. \textbf{The papers provide the first experimental demonstration of electronic correlation effects in flat bands of a \moire superlattice.}\\
Cao, Y. et al. Correlated insulator behaviour at half-filling in magic-angle graphene superlattices~\cite{cao_correlated_2018},\\
Cao, Y. et al. Unconventional superconductivity in magic-angle graphene superlattices~\cite{cao2018unconventional}.\\
5. \textbf{The paper reports the observation of anomalous Hall effect in a \moire superlattice.}\\
Sharpe, A. L. et al. Emergent ferromagnetism near three-quarters filling in twisted bilayer graphene~\cite{sharpe_emergent_2019}.\\
6. \textbf{The paper reports the observation of quantized anomalous Hall effect in a \moire superlattice}.\\
Serlin, M. et al. Intrinsic quantized anomalous hall effect in a moiré heterostructure~\cite{serlin2020intrinsic}.\\
7. \textbf{This review covers the different mechanisms responsible for anomalous Hall effect (AHE) observed in non-\moire materials.}\\
Nagaosa, N., Sinova, J., Onoda, S., MacDonald, A. H., Ong, N. P. Anomalous Hall effect~\cite{nagaosa_anomalous_2010}.\\
8. \textbf{We recommend this review on quantum Hall effect.}\\
von Klitzing, K. The quantized Hall effect~\cite{von_klitzing_quantized_1986}.\\
9. \textbf{We recommend this review for getting started on Berry physics ideas and its experimental effects.}\\
Xiao, D., Chang, M.-C., Niu, Q. Berry phase effects on electronic properties~\cite{xiao_berry_2010}.\\
10. \textbf{The paper provides a theoretical proposal of second harmonic Hall voltage generation due to Berry curvature dipole.}\\
Sodemann, I., Fu, L. Quantum nonlinear hall effect induced by berry curvature dipole in time-reversal invariant
materials~\cite{sodemann_PRL2015_quant}.\\
11. \textbf{An experimental work to detect topological transition in a \moire{} material, probed using Berry curavture dipole.}\\
Sinha, S. et al. Berry curvature dipole senses topological transition in a moiré superlattice~\cite{sinha_NP2022_berry}.\\
%12. \Moire{} probed using STM.\\
%Yankowitz, M. et al. Emergence of superlattice Dirac points in graphene on hexagonal boron nitride.\\
12. \textbf{A theoretical study on Berry phase inducing a valley-dependent Hall transport.}\\
Xiao, D., Yao, W., Niu, Q. Valley-Contrasting Physics in Graphene: Magnetic Moment and Topological Transport~\cite{xiao_valley-contrasting_2007}.\\
13. \textbf{This theoretical study highlights that \moire{} bands can be topological.}\\
Song, J. C. W., Samutpraphoot, P., Levitov, L. S. Topological Bloch bands in graphene superlattices~\cite{PNAS_song_topological_2015}.\\
14. \textbf{This review highlights the reproducibility issues and challenges in the fabrication of \moire superlattice devices.}\\
Lau, C. N., Bockrath, M. W., Mak, K. F., Zhang, F. Reproducibility in the fabrication and physics of moiré
materials~\cite{lau_reproducibility_2022}.\\
15. \textbf{This theoretical study highlights that TMDC \moire{} materials are a platform to realize correlated and topological states.}\\
Wu, F., Lovorn, T., Tutuc, E., Martin, I., MacDonald, A. Topological Insulators in Twisted Transition Metal Dichalcogenide Homobilayers~\cite{wu_topological_2019_2}.\\
16. \textbf{This experimental work is the first demonstration of valley Hall effect.}\\
Mak, K. F., McGill, K. L., Park, J., McEuen, P. L. The valley Hall effect in MoS$_2$ transistors~\cite{mak_valley_2014}\\
17. \textbf{This theoretical paper outlines the Haldane model that demonstrates the possibility of having quantum Hall effect in the absence of the magnetic field.}\\
Haldane, F. D. M. Model for a Quantum Hall Effect without Landau Levels: Condensed-Matter Realization of the "Parity Anomaly"~\cite{haldane_model_1988}.\\
18. \textbf{This paper reports the first experimental observation of quantum anomalous Hall effect in a topological insulator.}\\
Chang, C.-Z. et al. Experimental Observation of the Quantum Anomalous Hall Effect in a Magnetic Topological Insulator~\cite{chang_experimental_2013}.\\
19. \textbf{This theoretical paper proposes TMDCs as a Hubbard model simulator.}\\
Wu, F., Lovorn, T., Tutuc, E., MacDonald, A. Hubbard Model Physics in Transition Metal Dichalcogenide moiré Bands~\cite{wu_hubbard_2018}.\\
20. \textbf{Experimental demonstrations of fractional quantum anomalous Hall effect in \moire heterostructures.}\\
Cai, J. et al. Signatures of Fractional Quantum Anomalous Hall States in Twisted MoTe$_2$~\cite{cai_signatures_2023}.\\
Park, H. et al. Observation of Fractionally Quantized Anomalous Hall Effect~\cite{park_observation_2023}.\\
Zeng, Y. et al. Thermodynamic evidence of fractional Chern insulator in moiré MoTe$_2$~\cite{zeng2023thermodynamic}.\\
Xu, F. et al. Observation of integer and fractional quantum anomalous hall effects in twisted bilayer MoTe$_2$~\cite{xu_observation_2023}.\\
21. \textbf{First experimental report of the nonlinear Hall effect.}\\
Ma, Q. et al. Observation of the nonlinear Hall effect under time-reversal-symmetric conditions~\cite{ma_observation_2019}.\\
22. \textbf{These theoretical papers connect the idea of quantum metric to flat band superconductivity.}\\
Peotta, S., Törmä, P. Superfluidity in topologically nontrivial flat bands~\cite{Peotta2015}.\\
Julku, A., Peotta, S., Vanhala, T. I., Kim, D.-H., Törmä, P. Geometric origin of superfluidity in the lieb-lattice flat band~\cite{PhysRevLett.117.045303}.\\

\section*{Acknowledgements}
 We thank Joydip Sarkar, Ayshi Mukherjee, Krishnendu Maji, Surat Layek, Amit Basu, Chandni U., Biswajit Datta, Justin Song, and Jainendra K. Jain for the discussions. M.M.D. acknowledges the Department of Science and Technology (DST) of India for Nanomission grant SR/NM/NS45/2016 and DST SUPRA grant SPR/2019/001247 along with the Department of Atomic Energy of Government of India 12-R\&D-TFR-5.10-0100 for support.
A.A thanks the Department of Science and Technology for Project No. DST/NM/TUE/QM-6/2019(G)-IIT Kanpur, of the Government of India, for financial support.
P.C.A. acknowledges support by the National Science Foundation under Grant No. OMA-2328993.

\section*{Author contributions}
M.M.D. ideated and led the writing of this review. All authors discussed and contributed to the writing.

\section*{Competing interests}
The authors declare no competing interests.

%%________________________________________

\newpage 
\begin{tcolorbox}
\begin{center}
    {\textbf{Theory box}}
\end{center}
\footnotesize
{\it Origin of Berry curvature in transport:--}
The system's symmetry governs the flow of electrons in periodic solids, and the wavefunctions encode this information. In certain solids, the electron wavefunction has a "self-rotation" - which deflects the electron sideways in the presence of external fields. The Berry curvature physics captures this sideways deflection of electrons \cite{sundaram_PRB1999_wave, xiao_berry_2010}. We briefly summarize a formal way to understand this. 
A crystalline material interacting with an electromagnetic field ($\bm E$) can be described (within the length gauge) via the Hamiltonian, 
%\begin{equation}
$    {\cal H}= {\cal H}_{0}-\textit{e}~\hat{\boldsymbol r}\cdot{\boldsymbol E}\label{eq:1}~. 
$
%\end{equation}
Here, $e<0$ is the electronic charge, and $\hat{\bm r}$ is the position operator. 
 ${\cal H}_{0}$ is the unperturbed Hamiltonian with Bloch eigenstates,  
 ${\cal H}_{0}\ket{u_n(\boldsymbol{k})} = \varepsilon_{n}(\boldsymbol{k})\ket{u_n(\boldsymbol{k})}$, 
 specified by the band index $n$ and the crystal momentum $\boldsymbol{k}$. On including  the impact of the electric field perturbatively, we encounter the matrix elements of the $\hat{\bm{r}}$ operator, or the Berry connection 
 ${\bm {\mathcal R}}^a_{nm}= i \langle u_{n}({\bm k})| {\partial}_{k_a}  u_{m}({\bm k}) \rangle$.
 The Berry curvature arises from the product of two Berry connections, and it can be expressed as the commutator of the spatial coordinates, $[r_a, r_b] = i \epsilon_{abc} \Omega_{c}~$ \cite{ong-geomtery2005,zhou_berry_2015}. Here, $a,b,c$ denote the coordinate axes, and we have $r_a = {\mathcal R}^a_{pm}$ and $r_b = {\mathcal R}^b_{mp}$. 
 Physically, the Berry curvature captures the impact of inter-band coherence on transport and optical properties, and it satisfies the following symmetry properties, 
 \begin{equation}
 {\rm Time~reversal~symmetry :} ~~ {\bm \Omega}({\bm k})  \to -{\bm \Omega}({-\bm k})~,~~~~{\rm Space~inversion~symmetry:} ~~ {\bm \Omega}({\bm k})  \to {\bm \Omega}({-\bm k}) \nonumber~.
 \end{equation}
 As a result, if both time reversal and space inversion symmetry are protected, then $ {\bm \Omega}({\bm k})=0$ for all momentum. To realize a finite Berry curvature in the system, one has to break either inversion or time-reversal symmetry. 
      
\vspace{.2 cm}
The Berry curvature modifies the dynamics of the electron wavepacket in the presence of an external electric field. The rate change of position $\dot{\bm r}$ and momentum $\hbar\dot{\bm k}$ of the center of the mass of the electron wave 
packet in a given band is given by, 
%\begin{equation}\label{eom}
$\dot{\bm r} =   {\bm v}+\frac{e}{\hbar}({\bm E}\times {\bm \Omega}),~~{\rm and}~~\hbar\dot{\bm k} = -e{\bm E}~$.
%\end{equation}
Here, the band velocity is specified by $\hbar {\bm v}  =\nabla_{\bm k}{\varepsilon_{\bm k}}$, where  $\varepsilon_{\bm k}$ 
is the electronic dispersion. 
 The anomalous velocity ${\bm E} \times {\bm \Omega}$ gives rise to the anomalous Hall effect, and the Berry curvature dipole-induced nonlinear anomalous Hall effect~\cite{PhysRev.95.1154,nagaosa_anomalous_2010,sodemann_PRL2015_quant, sinha_NP2022_berry}. These interesting Hall responses are intricately related to the band topology of the system. For example, the Chern number of a given band in a time reversal symmetry broken two-dimensional system is given by $C = \sum_{\bm k} \Omega_z ({\bm k})$.
 
\vspace{.2 cm}
{\it Berry curvature and $B=0$ Hall effects:--} 
The current density $j_a$ induced by an electric field $E_b$ can be expressed as  
\begin{equation}\label{eq:secorderNLH}	
%\centering 
j_{a}=\sum_{b}\sigma_{ab}^{(1)}E_b+\sum_{bc}\sigma_{abc}^{(2)}E_bE_c+\sum_{bcd}\sigma_{abcd}^{(3)} E_bE_cE_d+...~, 
\end{equation}
%\end{center}
with ($a, b, c, d$) denoting the spatial coordinates, 
and $\sigma^{(n)}$ represents the conductivity tensor for the $n$-th order response. 
The symmetries of the system determine the leading order contribution to the current density.
The intrinsic anomalous Hall effect in magnetic systems and the valley Hall response in 2D hexagonal systems are prominent Hall responses induced by the Berry curvature in the linear response regime. At low temperatures, the anomalous Hall effect in magnetic materials probes the band topology of partially filled bands. 
The total anomalous charge Hall response vanishes in nonmagnetic systems like 2D hexagonal gapped graphene, which preserves time-reversal symmetry. Despite this, each of the system's two valleys can have equal and opposite anomalous Hall responses, which combine to generate a finite valley Hall effect \cite{polshyn2020electrical,deng_quantum_2020}.  
Additionally, if such nonmagnetic systems 
have broken inversion symmetry, the Berry curvature dipole can induce a dominant nonlinear (second order in the electric field) anomalous charge Hall response in them. 
The anomalous Hall, valley Hall and the nonlinear 
Hall responses are given by, 
 \begin{equation}
     \sigma_{ab}^{\rm{AHE}} = -\dfrac{e^2}{\hbar}\epsilon_{abc} \sum_{n,\bm k}\Omega^c_{n \bm{k}}f_{n\bm{k}}~;~~
     \sigma_{xy}^{\rm VHE} ({\rm total})= \sigma_{xy}^{ \rm AHE}(K) - \sigma_{xy}^{\rm AHE}(K')~;~~
    \sigma_{abc}^{\rm{BCD}} 
     = -\dfrac{e^3\tau}{2\hbar^2}\epsilon_{adc}\sum_{n \bm k}f_{n{\bm k}}\frac{\partial \Omega_{n\bm k}^d}{\partial k_b}~.
 \end{equation}
Here, $\epsilon_{abc}$ is the third rank antisymmetric tensor, $f_{n\bm{k}}$ is the equilibrium Fermi Dirac distribution function, and $K/K'$ specify the valley degree of freedom. 
The two valleys are time-reversed partners of each other. However, individual valleys break time-reversal symmetry, giving rise to a finite valley-specific Hall response, along with the valley Chern number.

\vspace{.2cm}
{\it B = 0 Hall effects in gapped graphene}:-- 
To demonstrate these Berry-curvature induced Hall responses, let's consider the example of gapped and tilted graphene. The adjoining figure highlights valley-contrasting 
\begin{wrapfigure}{r}{.3\linewidth}
\vspace{-7pt}
\begin{center}
\includegraphics[width=0.32\textwidth]{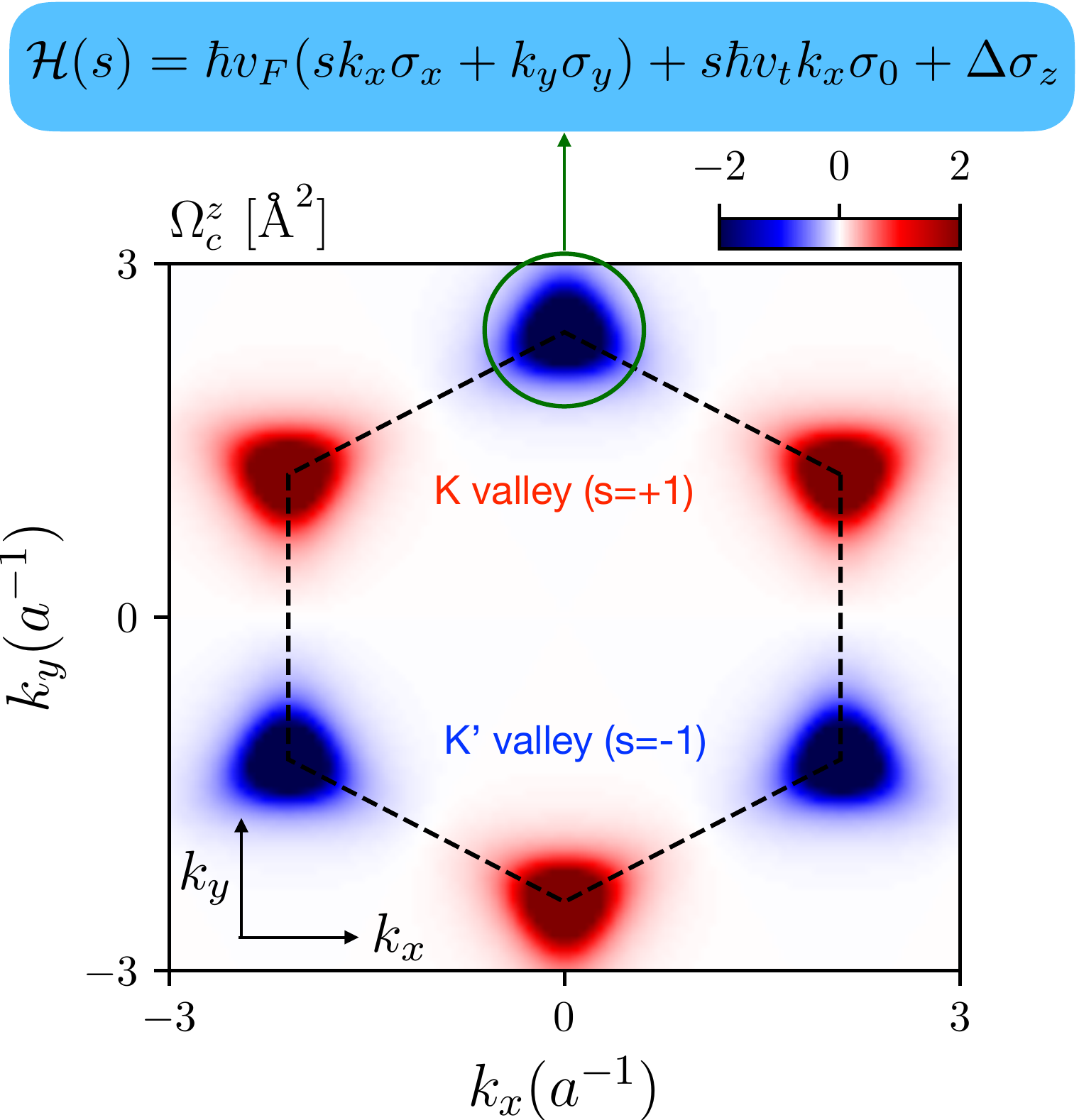}
\end{center}
\end{wrapfigure}
 Berry curvature %($z$-component) 
of the inversion symmetry broken hexagonal graphene lattice and the low energy Hamiltonian for momentum points near the valleys. The tilt velocity $v_t$ in the Hamiltonian is typically induced by strain, which breaks crystalline symmetries and makes the bands and the Fermi velocity anisotropic. 
  The valley contrasting Berry curvature and the anomalous Hall effect for each valley are
 \begin{equation}
     \Omega^z(s) = s \dfrac{\hbar^2v_F^2\Delta}{2(\hbar^2 v_F^2k^2+\Delta^2)^{3/2}}~;~~~~\sigma_{xy}^{\rm{AHE}}(s) = -s\dfrac{e^2}{2\hbar}\biggl(1-\dfrac{\Delta}{\mu}\biggr)~,
\end{equation}
with $\mu>0$ being the chemical potential, and $s=\pm1$ denotes the two valley's. 
While the system's total anomalous Hall effect vanishes if time-reversal symmetry is preserved,  
 the valley Hall and the Berry curvature dipole induced nonlinear Hall conductivity are finite. These are given by, 
\begin{equation}
\sigma_{xy}^{\rm VHE} ({\rm total})= - \frac{e^2}{\hbar}\biggl(1-\dfrac{\Delta}{\mu}\biggr)~,~~{\rm and}~~
   \sigma_{yxx}^{\rm{BCD}} = v_t~\dfrac{e^3\tau}{\hbar^2}\dfrac{3\Delta}{16\pi \mu^2}\left[1-\biggl(\dfrac{\Delta}{\mu}\biggr)^2 \right].
\end{equation}
For completely filled conduction band ($\mu \gg \Delta$), the valley Hall effect becomes quantized in units of $e^2/\hbar$. The nonlinear Hall conductivity is $\propto v_t$; this highlights the crucial role that breaking crystalline symmetries and an anisotropic band dispersion 
play in non-linear Hall response.

 \end{tcolorbox}

\clearpage

%------------------------------------------------------------------------
\clearpage
%----------------------------------------------------------------------

%------------------------------------------------------------------------
%----------------------------------------------------------------------

%------------------------------------------------------------------------
%----------------------------------------------------------------------

%-----------------------------------------

\end{document}